\documentclass[useAMS,usenatbib]{mn2e}

\usepackage{amssymb}
\usepackage{graphicx}
\usepackage[usenames,dvipsnames]{color}
\usepackage{verbatim}

\usepackage{times}
\usepackage{mathptmx}

\newcommand{\um}[1]{\,\mathrm{#1}}
\newcommand{\mic}[1]{\,\umu\mathrm{#1}}

\title[OH masers toward MIPSGAL bubbles]{Searching for OH maser emission towards the MIPSGAL compact Galactic bubbles}
\author[A. Ingallinera et al.]{{\large A. Ingallinera$^1$\thanks{E-mail:ingallinera@oact.inaf.it}, C. Trigilio$^1$, P. Leto$^1$, G. Umana$^1$, C. Buemi$^1$, L. Cerrigone$^2$, C. Agliozzo$^{3,4}$}\\
$^1$INAF-Osservatorio Astrofisico di Catania, Via Santa Sofia 78, 95123 Catania, Italy\\
$^2$ASTRON, the Netherlands Institute for Radioastronomy, PO Box 2, 7990 AA Dwingeloo, The Netherlands\\
$^3$Millennium Institute of Astrophysics, Santiago, Chile\\
$^4$Universidad Andres Bello, Avda. Republica 252, Santiago, Chile
}

\begin{document}

\date{Accepted 2015 August 12.  Received 2015 August 06; in original form 2015 July 20.}

\pagerange{\pageref{firstpage}--\pageref{lastpage}} \pubyear{2002}

\maketitle

\label{firstpage}

\begin{abstract}
We conducted radio observations searching for OH 18-cm maser emission from a sample of 169 unclassified MIPSGAL compact Galactic bubbles. These sources are thought to be the circumstellar envelopes of different kinds of evolved stars. Our observations were aimed at shedding light on the nature of MIPSGAL bubbles, since their characterisation is a fundamental aid for the development of accurate physical models of stellar and Galaxy evolution. The maser emission is observatively linked to the last stages of the life of low- and intermediate-mass stars, which may constitute a significant fraction of the MIPSGAL bubbles. In particular OH masers are usually observed towards post-AGB stars. Our observations were performed with the Green Bank Telescope and, for each source, produced spectra around the four OH 18-cm transitions. The observations were compared with archive interferometer data in order to exclude possible contamination from nearby sources. The main result is that the OH maser emission is not a common feature among the MIPSGAL bubbles, with only one certain detection. We conclude that among the MIPSGAL bubbles the post-AGB stars could be very rare.
\end{abstract}


\section{Introduction}
At the end of their life, low- and intermediate-mass stars ($\la8\um{M}_\odot$ at the beginning of main-sequence) evolve from the asymptotic giant branch (AGB) towards the planetary nebula (PN) phase. During the AGB period, the star undergoes a relevant mass loss \citep{Habing1996} and sheds its outer layers, which form a circumstellar envelope (CSE). As the CSE expands, it gets cooler and the material can condense to form dust. The CSE absorbs the radiation from the central star and re-emit the absorbed light in the infrared (IR). The AGB phase definitely ends when most of the outer layers have been shed. Then, the net mass-loss rate decreases and the stellar wind becomes hotter and faster. This post-AGB stage represents the intermediate phase between AGB stars and PNe \citep{vanWinckel2003}. In this phase the fast and hot wind sweeps up the stellar ejecta to form circumstellar shells. The resulting dust and gas nebula surrounding the post-AGB stars is usually referred to as protoplanetary nebula (PPN)\footnote{Indeed, the terms `post-AGB' and `PPN' are usually treated as synonyms.}. This process ends when the central star, having lost entirely its outer layers, exposes its inner layers, hot enough to ionise the CSE. Finally the star enters the PN phase.

During the AGB phase the stellar wind is characterised by a general isotropy, responsible for the spherical symmetry of the CSE. However more than half PNe show elliptical or bipolar shapes, and similar asymmetric shapes are observed in PPNe. It was thought that this loss of symmetry arose during the early post-AGB phase (see \citealt{Sahai1998} and references therein), however recent studies show that the stellar wind can undergo a departure from spherical symmetry already in the AGB phase \citep{Ragland2008}. On the other hand, roundish PNe reasonably derive from roundish PPNe, where a roughly spherical symmetry has never been lost (see \citealt{Ramos2012} for a discussion about round PPN and PN candidates). The study of the different morphologies of PPNe has become fundamental to develop accurate magnetohydrodynamics models of the evolution of the CSE \citep{Velazquez2014}, which, in turn, influences the chemical enrichment of the interstellar medium. Therefore the identification of new PPNe provides remarkable insights and constraints on the study of the fundamental physics involved in the last stages of stellar life and in the interstellar medium evolution.

In two previous works \citep{Ingallinera2014a,Ingallinera2014b} we reported the open issue of the classification and characterisation of the compact Galactic bubbles (`MBs' hereafter; \citealt{Mizuno2010}) discovered with the \textit{Spitzer Space Telescope} at $24\mic{m}$ by the MIPSGAL survey \citep{Carey2009}. In particular, we showed how radio observations represent a valid and valuable instrument for this purpose. The MBs, at $24\mic{m}$, are characterised by an apparent roundish morphology, small angular dimensions (usually $<1\um{arcmin}$) and a usual lack of counterpart at other wavelengths. They are thought to be the circumstellar shells of different kinds of evolved stars. Currently about 30 percent of the MBs are classified, $\sim\!45$ percent of them as PNe, but with examples of massive stars (luminous blue variables, Wolf--Rayet stars, G and K giants, OB stars) and supernova remnants \citep{Flagey2014} as well. Since among the classified object one is a PPN (see Section \ref{sec:spitzer}), it is not possible to rule out the possibility that some other MBs could be PPNe.

In this paper we briefly discuss about the infrared morphology of known example of PPNe and we infer which would be the most likely expected morphology for a PPN in the MBs sample (Section \ref{sec:spitzer}). As an independent test to search for possible PPNe among the MBs, in Section \ref{sec:OH} we report new observations searching for OH masers towards a sub-sample of 169 MBs. OH masers in fact are a common feature of Oxygen-rich AGB and post-AGB stars, observed in about 80 percent of them. Indeed the presence of OH maser emission during the whole post-AGB phase, or even later, is associated with asymmetric stellar winds and CSEs \citep{Cerrigone2013}. For spherical symmetric post-AGB stars OH masers should be observed only in their early moments ($\sim\!1000\um{y}$; \citealt{Lewis1989}). It is important to notice that many bubbles are only barely resolved in MIPSGAL and therefore a roughly circular appearance may hide an actual asymmetric shape. For these reasons it is however worth to investigate the possible OH maser emission in the MBs.

\section{IR morphology of post-AGB stars}
\label{sec:ir}
Since the post-AGB phase is relatively short ($\la10^4\um{y}$), these objects are quite rare. Indeed, the first systematic searches for post-AGB stars became feasible only with the IR satellites. The heated dust envelope makes the post-AGB stars have very peculiar IR colours. In fact, in the last three decades, one of the most reliable way to identify AGB and post-AGB stars has been through their \textit{IRAS}\footnote{InfraRed Astronomical Satellite.} colour (e.g. in \citealt{Volk1989}). However, the resolution of \textit{IRAS} did not allow any morphological study of their CSE, and all they appeared as point sources.

Ground- and space-based instruments, especially in the last years, are efficiently improving and complementing \textit{IRAS} observations of AGB and post-AGB stars. For example, \citet{Lagadec2011} carried out mid-IR observations with the Very Large Telescope and the Gemini Observatory aimed at studying the dust distribution of 93 objects (AGB and post-AGB stars, young PNe and massive evolved stars). They were able to resolve 17 PPNe, showing the different possible shapes that these objects present. The morphology of two of them, namely IRAS 07134+1005 and IRAS 19500-1709, strikingly resemble that of MBs. Many others instead present a relevant asymmetry. The typical angular dimension of the PPN reported in their work is less than $10\um{arcsec}$, considerably smaller than the typical MBs extension. These observations are very important because they point out that: a) in the mid-IR the PPN is not necessarily heavily outshined by the central post-AGB stars; b) many PPNe could be only barely resolved in MIPSGAL (resolution $\sim\!6\um{arcsec}$) and all the different shapes can result in a unique roundish appearance. Therefore it is, at least plausible, that several PPNe can be present in the MBs sample.

\subsection{Post-AGB stars and PPNe with \textit{Spitzer}}
\label{sec:spitzer}
The improvement in sensitivity and resolution introduced by \textit{Spitzer} gave the possibility of a space-based imaging for this class of stars. In fact, at least for near AGB and post-AGB stars, it would have been possible to resolve their CSE.

To have a first clue on the typical PPN appearance in MIPSGAL, we took the post-AGB catalogue by \citet{Deacon2004} and the Toru\'n catalogue \citep{Szczerba2007} to search for post-AGB stars in the survey tiles. We found that these objects appear as extremely bright point sources and about 90 percent of them produces a saturated image. Furthermore their high brightness translates to evident sidelobes that make it hard to investigate their CSE.

\begin{figure}
\begin{center}
\includegraphics[width=\columnwidth]{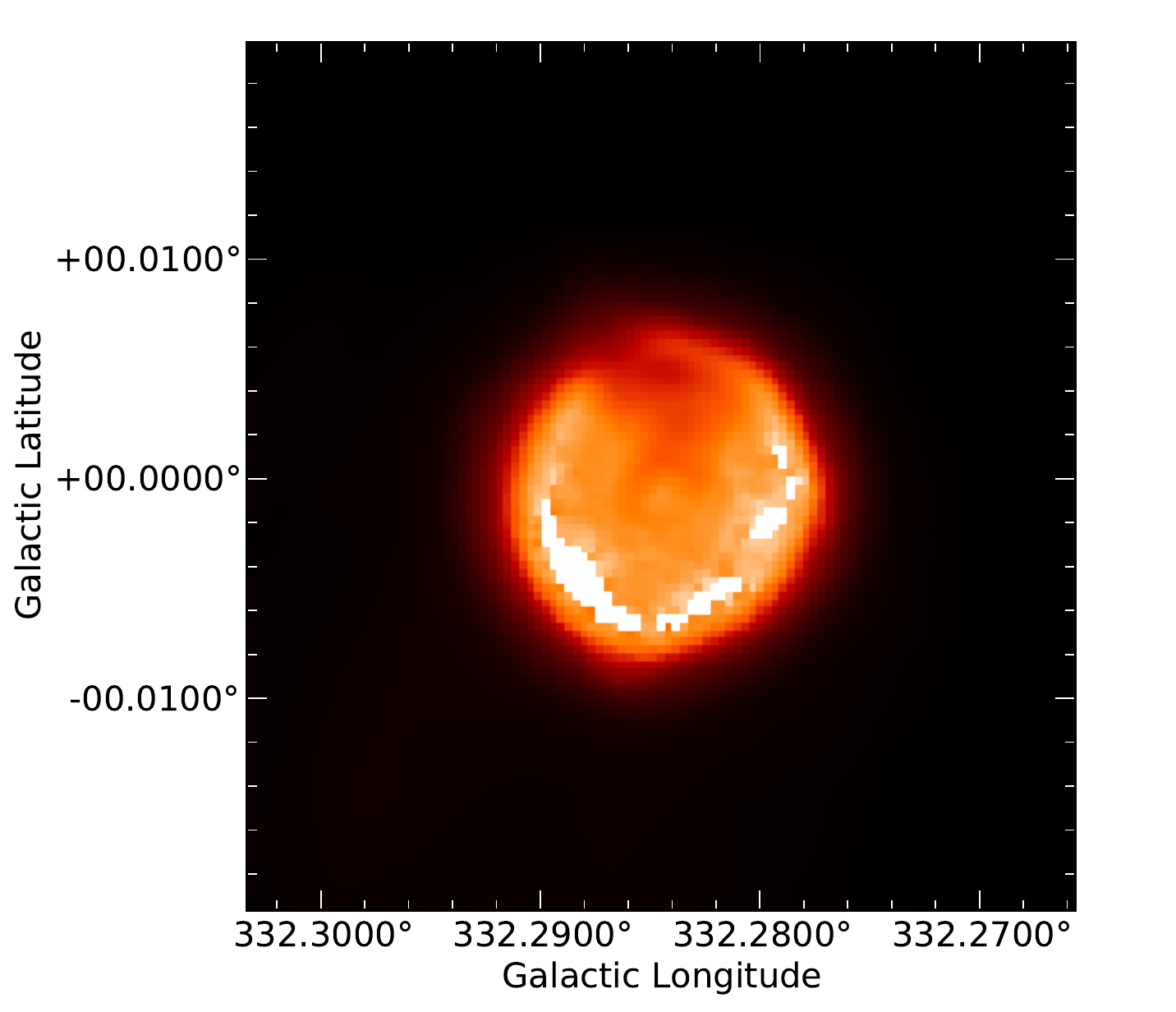}
\caption{MIPSGAL image at $24\mic{m}$ centred on the PPN IRAS 16115-5044. The southern region of the nebula is saturated.}
\label{fig:PPN_M24}
\end{center}
\end{figure}

Only one PPN of the Toru\'n catalogue was clearly resolved in MIPSGAL, namely the PPN IRAS 16115-5044. This source is indeed associated with the MB MGE 332.2843-00.0002, and it is the only known PPN among the MBs. At $24\mic{m}$, following the \citet{Mizuno2010} morphological classification, it appears as a very bright roundish source with a detected central source (Figure \ref{fig:PPN_M24}) and a diameter of $47\um{arcsec}$.
The central source is still visible at least down to $1.25\mic{m}$ (2MASS $J$-band), while there is no evidence of the CSE at wavelengths shorter than $24\mic{m}$. \citet{Weldrake2003} estimate a distance of $2\um{kpc}$, which would translate in a CSE diameter of about $0.5\um{pc}$.

On the other hand, about 95 percent of MBs classified as PN present a `disk' or a `ring' shape at $24\mic{m}$, without a central source, neither at shorter wavelengths. The terms `disk' and `ring' describe only the observation appearance of the MBs (not their intrinsic morphology), the former is referred to round sources with a relatively flat brightness profile while the latter to round sources whose brightness significantly increases toward the edge (please refer to \citealt{Mizuno2010} for the complete nomenclature). Their typical diameter is usually around $20\um{arcsec}$, about a half of IRAS 16115-5044.

\begin{figure}
\begin{center}
\includegraphics[width=\columnwidth]{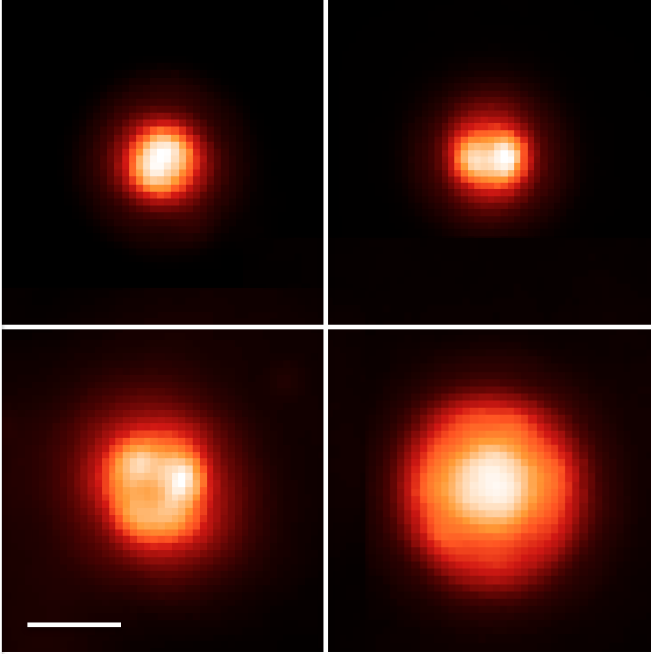}
\caption{Examples of PNe among the MBs (clockwise from upper-left: MGE~001.0178-01.9642, MGE~003.5533-02.4421, MGE~040.3704-00.4750, MGE~031.9075-00.3087). The white rule is 15-arcsec long.}
\label{fig:pn}
\end{center}
\end{figure}

\begin{figure}
\begin{center}
\includegraphics[width=6cm]{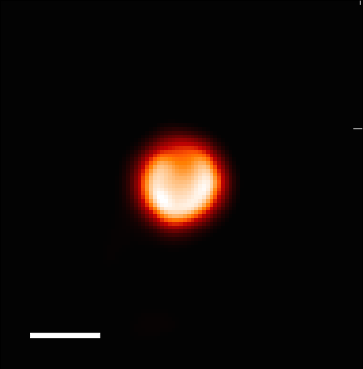}
\caption{Simulated image of the PPN IRAS 16115-5044 at $24\mic{m}$ posed twice as farther as it is. The white rule is 15-arcsec long.}
\label{fig:PPN_far}
\end{center}
\end{figure}

\begin{figure}
\begin{center}
\includegraphics[width=\columnwidth]{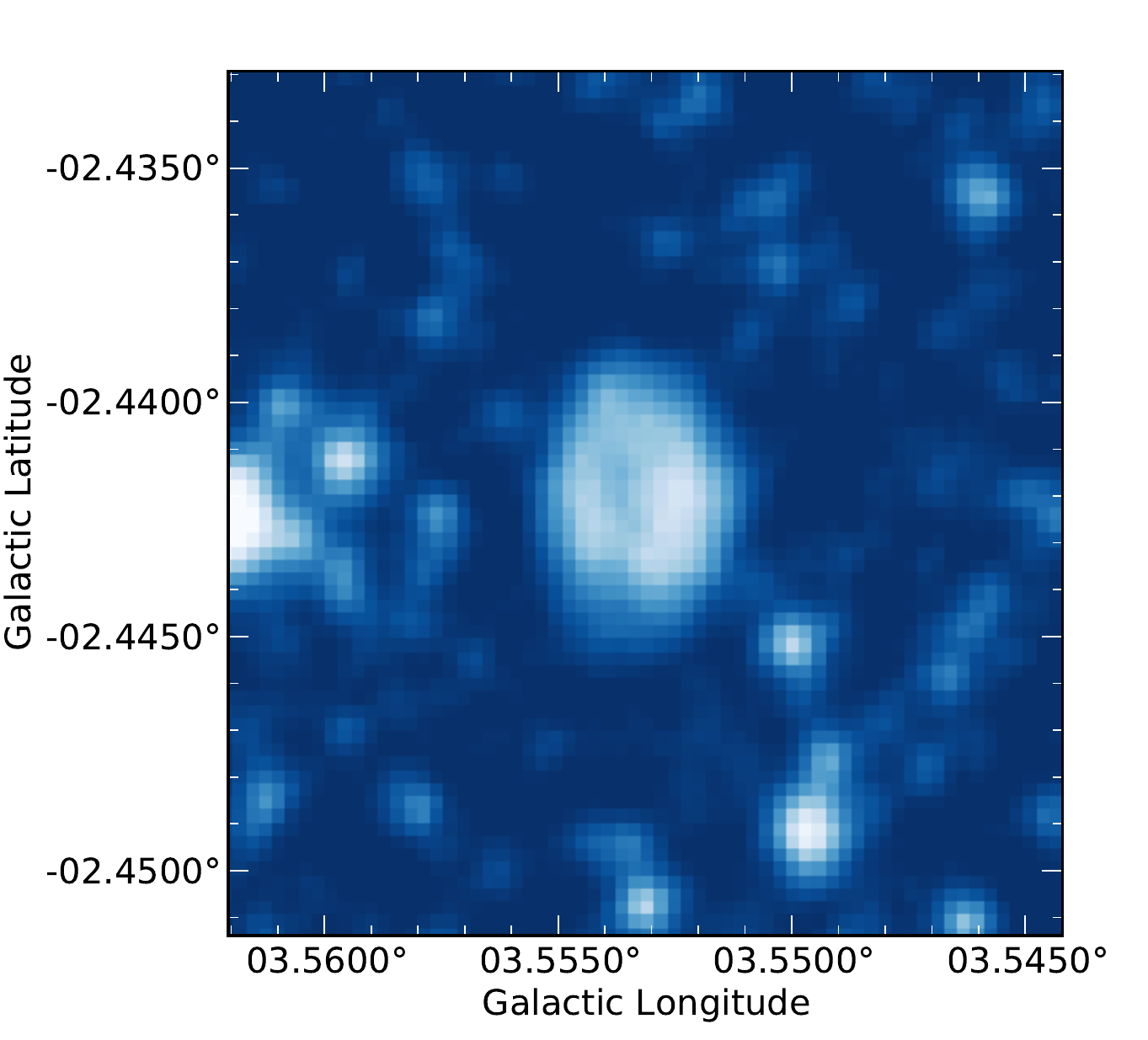}
\caption{Optical image of MGE~003.5533-02.4421 from the STScI Digitalized Sky Survey.}
\label{fig:PN2_DSS}
\end{center}
\end{figure}

The examples presented in this Section are useful to determine how a potential PPN among the MBs would appear. We have showed that at $24\mic{m}$ a large sample of known post-AGB stars appear as very bright point sources. On the other hand, almost all the MBs classified as PN are `disks' or `rings' (Figure \ref{fig:pn}; \citealt{Nowak2014}).
We infer that the most probable morphology for a PPN MB would be that of a `disk'. In fact the `disk' shape can be viewed as produced in several possible ways: a) the source is a filled shell; b) the source is an empty shell but it is too compact or too far to be resolved as a `ring'; c) there is still a central source visible at $24\mic{m}$ whose brightness is comparable with that of its `ring'-shaped CSE, and it completely fills the `ring' hole; d) the source is intrinsically asymmetric but is only barely resolved; e) the source is intrinsically asymmetric but is oriented in a such a way to appear roundish. To support this conjecture, in Figure \ref{fig:PPN_far} we report a simulated view of IRAS 16115-5044 as it would appear if it were twice as farther.
It is possible to notice a clear emerging `disk' appearance in agreement with our theoretical speculation. On the contrary in Figure \ref{fig:PN2_DSS} we report a higher-resolution optical image of PN MGE~003.5533-02.4421 (known also as IC 4673) whose MIPSGAL image has been showed in Figure \ref{fig:pn}. We can see how a clear non-spherical morphology ($22''\times15''$ as calculated by \citealt{Tylenda2003}) can be easily confused with an almost perfect round shape when observed at lower resolution or at different wavelengths.

\section{Searching for OH masers}
\label{sec:OH}
According to what has been stated in the last section, the presence of a relevant number of PPN among the MBs was a plausible hypothesis that deserved further considerations and an independent check. A very important aid came from radio observations, searching for maser emission. In fact, in the CSE of AGB and post-AGB stars, conditions for maser emission from different molecules occur. The presence of OH, SiO and H$_2$O masers is a common feature of the CSE of this kind of stars, and their appearance and disappearance has been used as a powerful indicator of their evolutionary state \citep{Lewis1989}. Maser emission from OH molecules strongly characterises PPNe, with detection rate around 80 percent (e.g. \citealt{Szymczak2004}). Therefore, if a relevant number of PPNe were present in the MBs then we should be able to detect maser emission from many of them.

Among the possible choices, we decided to observe the four OH 18-cm transitions. These lines arise from ground state of OH ($^2\Pi_{3/2}$, $J=3/2$), split by $\Lambda$ doubling and hyperfine interaction into four sublevels \citep{Dawson2014}. The rest frequencies of the resulting transitions are 1612.231, 1665.402, 1667.359 and $1720.530\um{MHz}$ (for simplicity we will refer to them dropping the decimal digits). The lines forming at 1665 and $1667\um{MHz}$ are usually referred to as the `main lines', while the others as the `satellite lines'. The choice to observe at this frequency was motivated by the fact that OH maser emission at 1612, 1665 and $1667\um{MHz}$ are observed for most of the duration of the post-AGB phase, unlike SiO and H$_2$O (the $1720\um{MHz}$ masers characterise shocked nebulae, like SNRs, instead). Furthermore it is easy to observe simultaneously all the four transitions given their small frequency separation. This possibility is fundamental to identify the emission arising from the diffuse interstellar environment, rather than from single sources (see Section \ref{sec:res}).

\subsection{Sample selection and instrument settings}
The observed sources are a sub-set of the total 428 MBs. In particular we first extracted those objects with a declination $\delta>-30^\circ$, well suited for observations from the northern hemisphere. Subsequently we excluded the sources already classified in SIMBAD: none of them, in fact, is classified as PPN and the very first general goal of the study of the MBs is their classification. The final sample contained 169 sources. We did not pose any constraint deriving from their 24-$\umu$m shape. For checking purpose we observed the well known maser sources OH26.5, VX Sgr and W51M. 

A pilot experiment, on a smaller sample of sources, was carried out with the Medicina 32-m radio telescope on July 2011. However, because of a severe contamination by radio-frequency interferences (RFI), line detection proved almost impossible. We therefore decided to use the Green Bank Telescope (GBT; project 12A-106), given its extremely sensitivity and its location in a radio quiet zone. Furthermore the higher resolution achievable with the GBT, with respect to Medicina radio telescope, would have helped us to reduce spurious detections (see Section \ref{sec:res}).

The observations with the GBT were performed in four different blocks between January and March 2012, for a total amount of 22 hours. Each source was observed once (mostly), twice or three times, in frequency switching mode. Since the observations could be regarded as `blind', the number of different scans per source was driven only by a minimisation of the telescope slewing time. With this configuration, the total on/off duration of each scan was $4\um{min}$, split in 120 1-s long acquisitions on and 120 off.

As front-end we used the $L$-band receiver in Gregorian focus, with a system temperature $\sim\!20\um{K}$ and a gain of $2.0\um{K\ Jy}^{-1}$. We observed in 4 spectral windows (each spectral window consists of one `on' and one `off' sub-band) by means of the GBT Spectrometer, in double polarisation mode (circular polarisation). The spectral windows were chosen so as to observe all the four lines associated with the OH 18-cm transitions. The bandwidth for each spectral window was $12.5\um{MHz}$ with a channel spacing of $3\um{kHz}$, allowing us to reach a velocity resolution of about $0.5\um{km\,s}^{-1}$. The expected rms was of order of $10\um{mJy}$.

\subsection{Data reduction}
A preliminary data reduction was executed by means of the \textsc{gbtidl} package. The raw data were calibrated in flux density and reported in kinematic LSR velocity.

For each spectral window we had two sub-bands (one `on' and one `off') observed in two polarisations (acquired simultaneously), for a total of 4 scans per spectral window per source. Each scan consisted of 120 different 1-second acquisitions, so that it could be considered as a $120\times4096$ matrix, where 4096 is the number of channels. Mapping this matrix is extremely useful to detect and flag RFI-corrupted data. In fact, interferences usually appear as spikes well confined in time. Once the degraded acquisitions have been dropped, the four different scans for each spectral window were averaged together in order to obtain a mean spectrum. Each mean spectrum was corrected for bandpass gain by fitting an opportune polynomial. Finally the fitted polynomial curve was regarded as an estimate of the continuum background and subtracted from the mean spectrum.

At the end of these process a continuum-subtracted spectrum for each scan and then for each spectral window was produced.

\subsection{Results}
\label{sec:res}
\begin{figure*}
\begin{center}
\includegraphics[width=7.5cm]{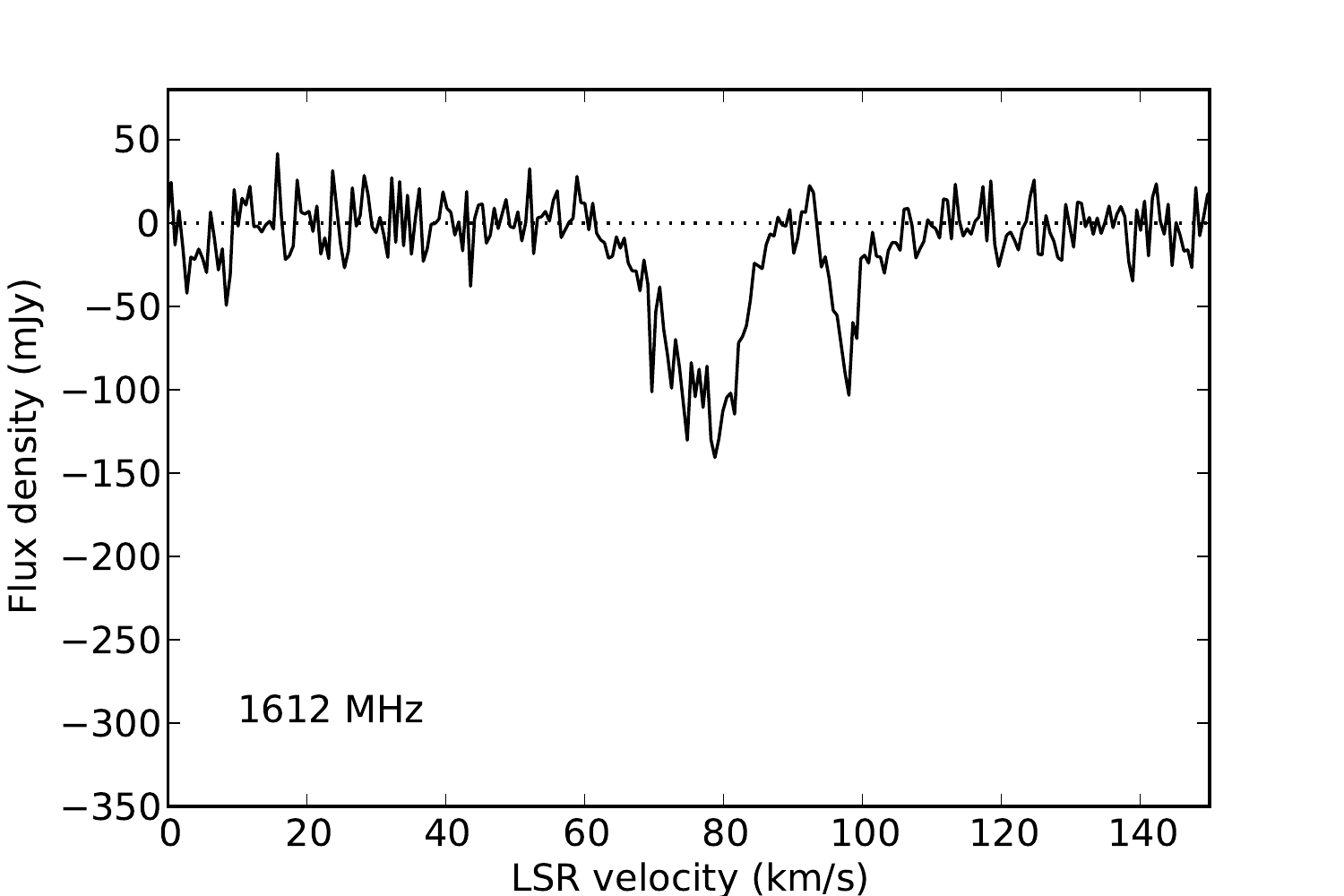}
\includegraphics[width=7.5cm]{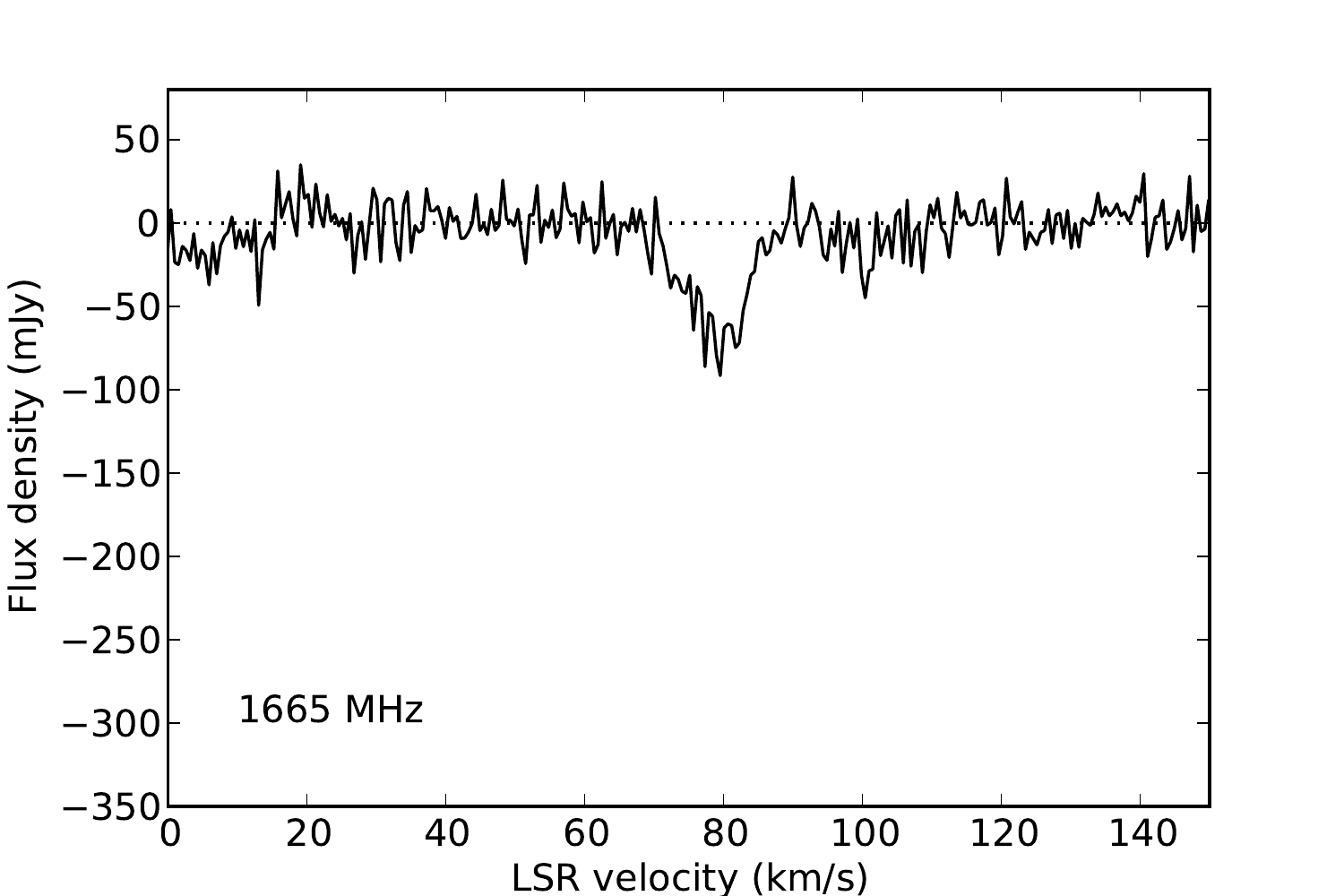}
\includegraphics[width=7.5cm]{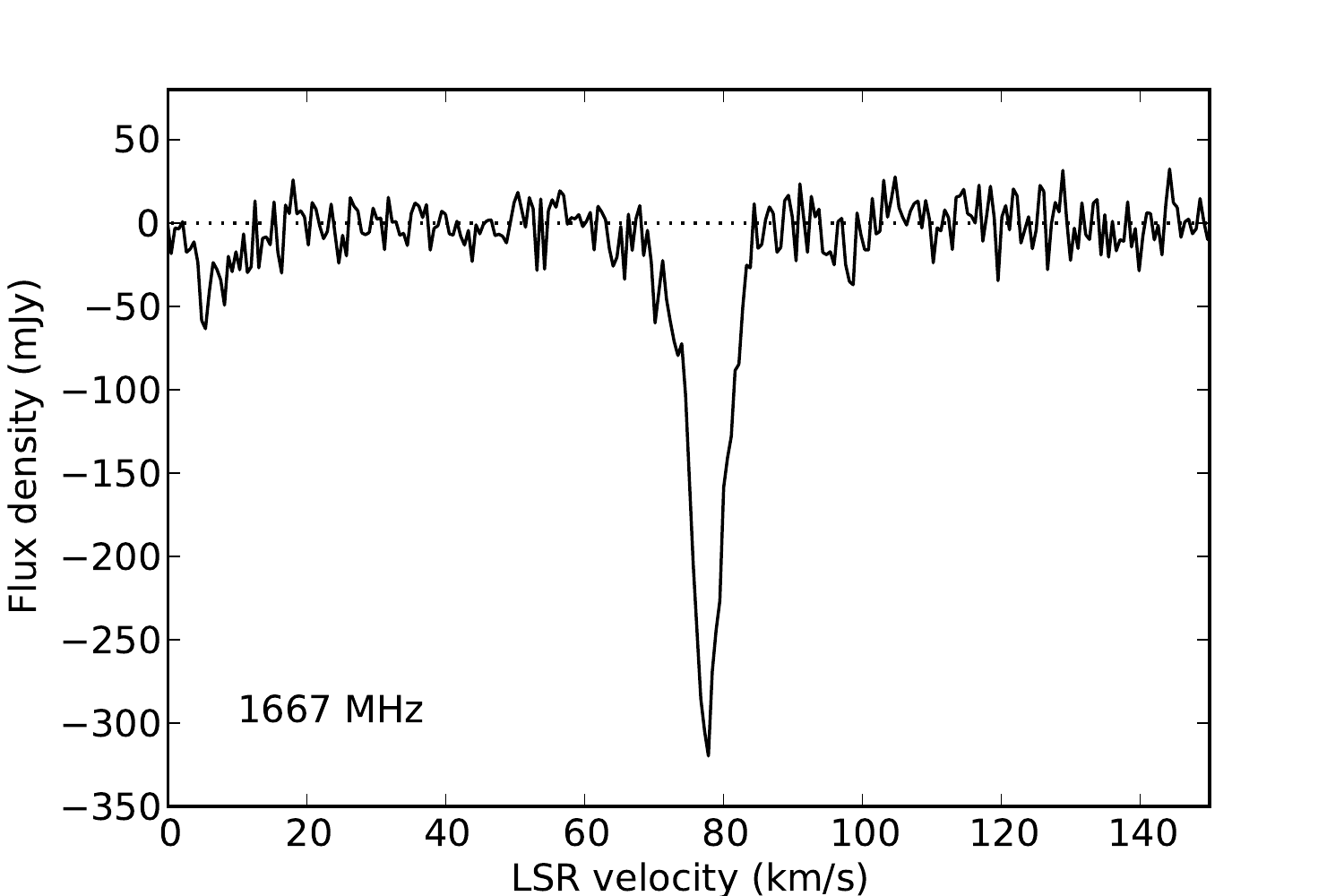}
\includegraphics[width=7.5cm]{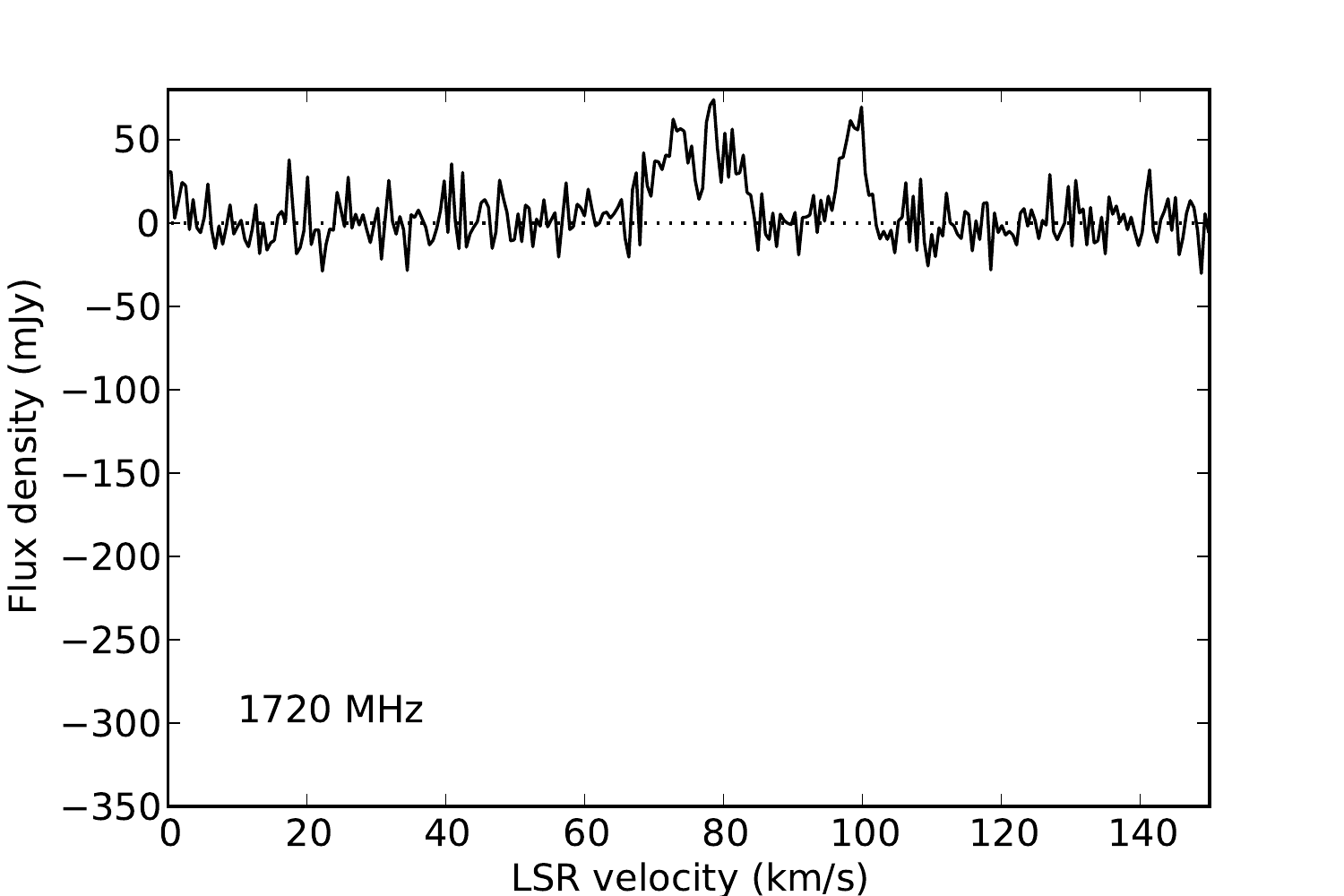}
\caption{Spectrum toward the source MGE 028.4451+00.3094 at 1612 (upper left), 1665 (upper right), 1667 (lower left) and $1720\um{MHz}$ (lower right), highlighting the diffuse interstellar medium signature. The two lines have a velocity of about 75 and $100\um{km\ s}^{-1}$, with the second barely detected in the main lines. We can notice also the departure from the 5:9 ratio expected for the main lines in case of LTE.}
\label{fig:diffuse}
\end{center}
\end{figure*}

Among the 169 spectra, 139 show OH lines, in emission or in absorption, in at least one spectral window. In 109 we find the peculiar features of the diffuse interstellar medium: the main lines appear in absorption with a height ratio close to 5:9 (expected in case of LTE); the two satellite lines show the same height but one in emission and the other in absorption \citep{Dawson2014}. In this case the four lines have approximately the same profile, the same velocity and the same width. They usually appear as broad lines, whose width ranges from $\sim\!10\um{km\ s}^{-1}$ to $\sim\!100\um{km\ s}^{-1}$. A typical spectrum with the diffuse medium signature is reported in Figure \ref{fig:diffuse}. The recognition of these patterns allowed us to exclude these lines from the maser search.

We ended up with 44 spectra showing emission features that could not be ascribed only to diffuse medium. Five spectra show multiple lines, likely emerging from different sources within the GBT beam (about $7.6\um{arcmin}$ FWHM at $1666\um{MHz}$). Following the notation introduced by \citet{Deacon2004}, we found 44 double-peaked (D-type) and 7 single-peaked (S-type) lines at $1612\um{MHz}$ and one D-type line at $1667\um{MHz}$. We were conscious that because of the large GBT beam (with respect to an interferometer) and because of its high sensitivity many detected masers could have been located a few arcminutes away from the pointing centre. As first assessment, we checked the catalogue of the ATCA/VLA OH $1612\um{MHz}$ survey by \citet{Sevenster1997a,Sevenster1997b,Sevenster2001}. The major clue to cross identify known masers was to compare line velocities. We noticed that many masers could be identified with sources very distant from the pointing centre: we measured distances up to half a degree. Some sources were then detected through GBT sidelobes. The sensitivity of the GBT and the very high brightness of maser emission are the causes of this issue. In fact the relative heights of the first two sidelobes (at $\sim\!17\um{arcmin}$ and $\sim\!24\um{arcmin}$ from the pointing centre) are approximately $\sim\!1/900$ and $\sim\!1/3000$ (see \citealt{Boothroyd2011} for the GBT beam model). Given these values, a maser line with a brightness of tens or hundreds jansky would be still easily detected with our sensitivity.

\begin{figure*}
\begin{center}
\includegraphics[width=6cm]{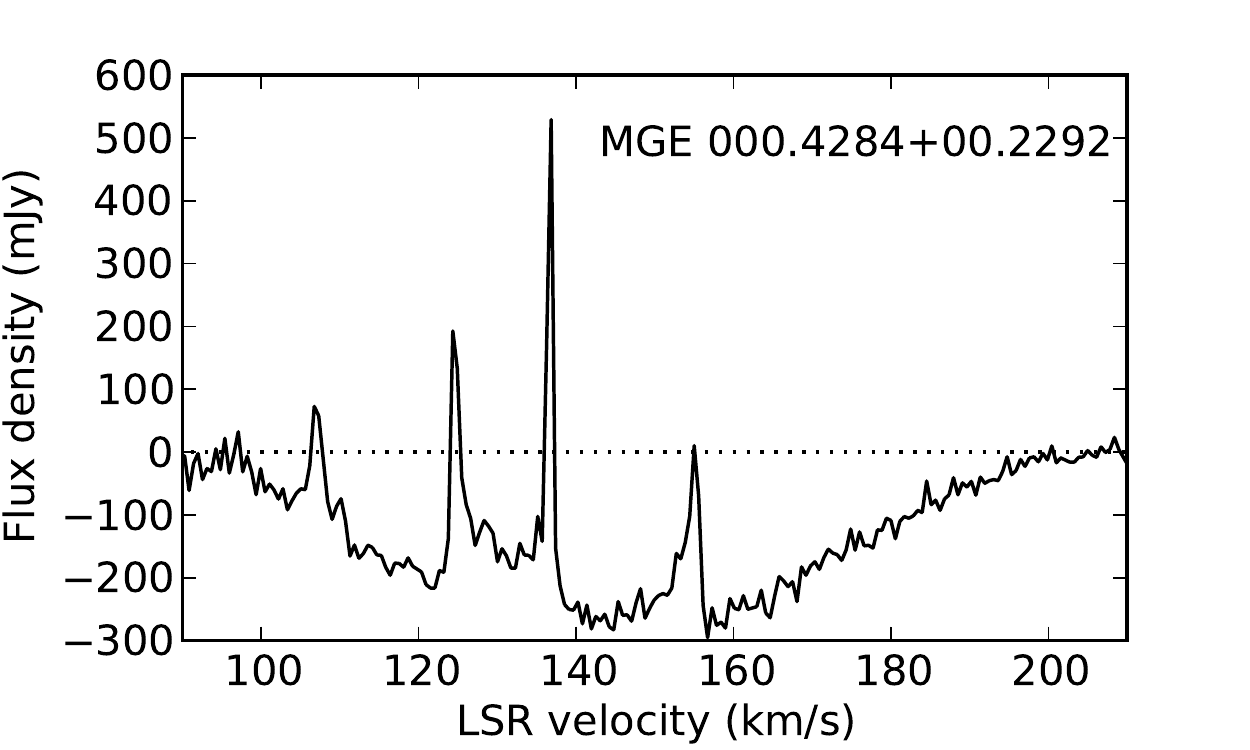}\hspace{-0.25cm}
\includegraphics[width=6cm]{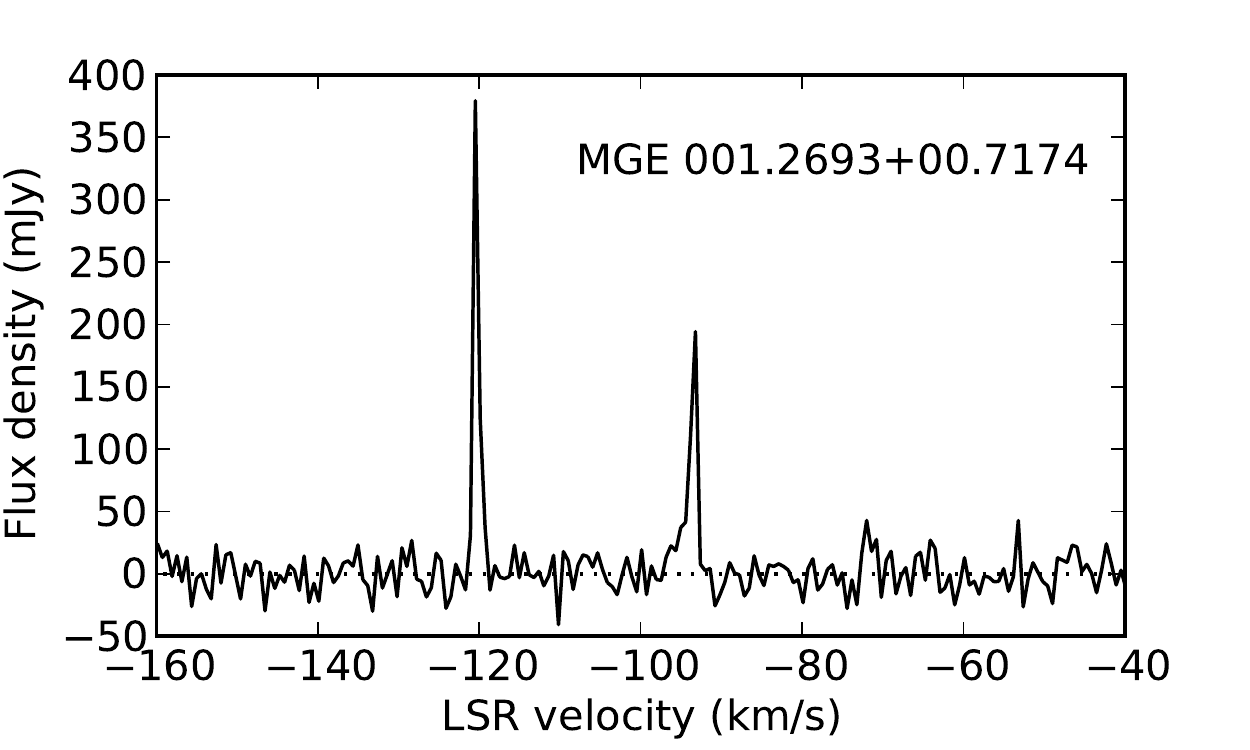}\hspace{-0.25cm}
\includegraphics[width=6cm]{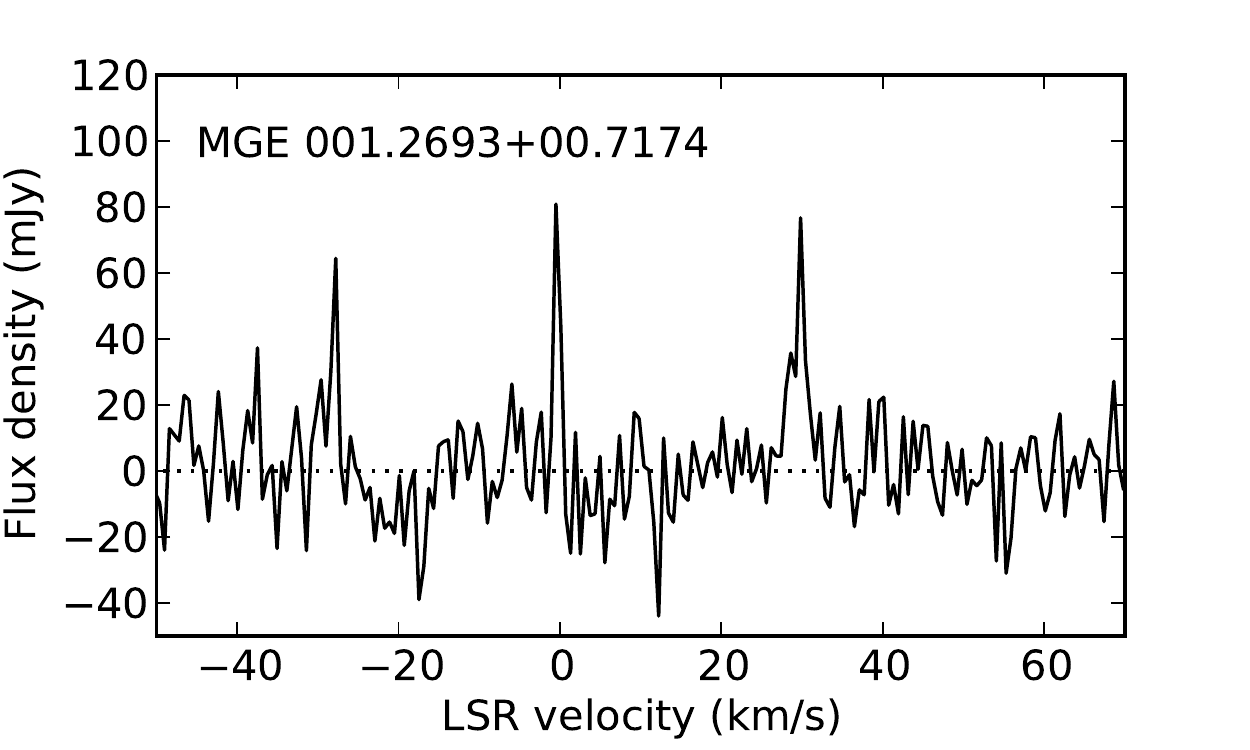}
\includegraphics[width=6cm]{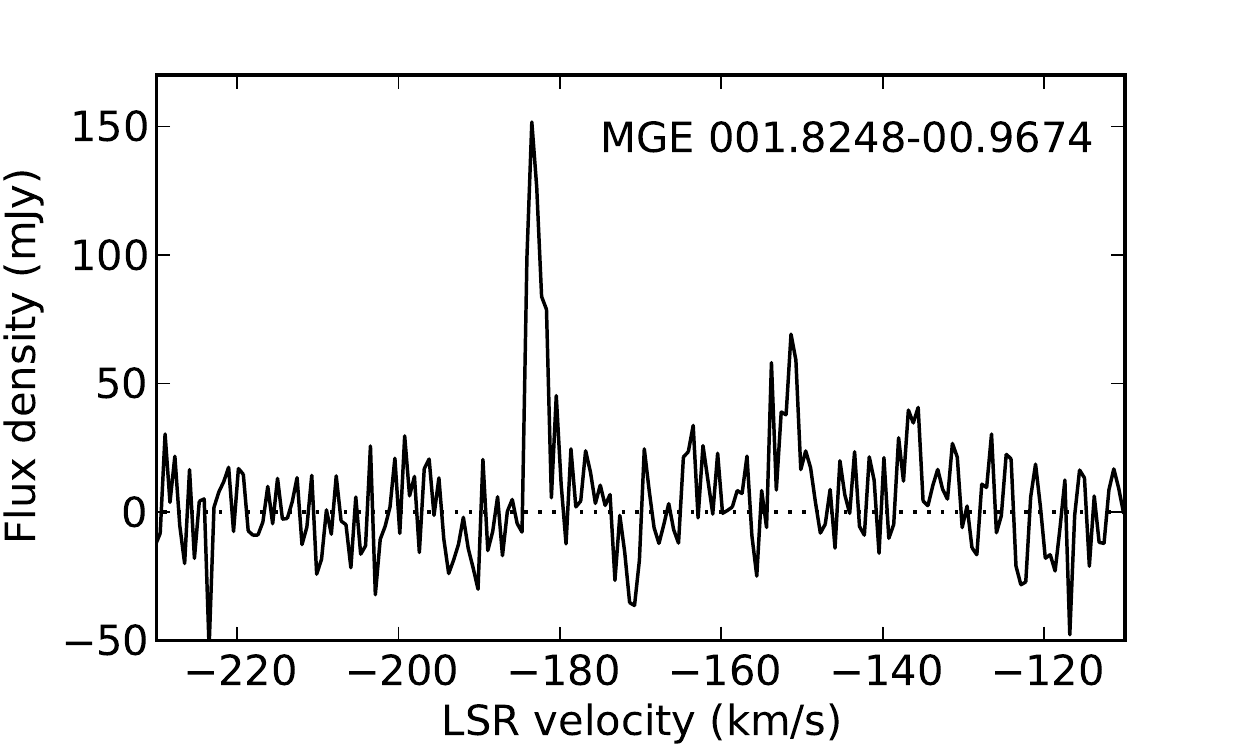}\hspace{-0.25cm}
\includegraphics[width=6cm]{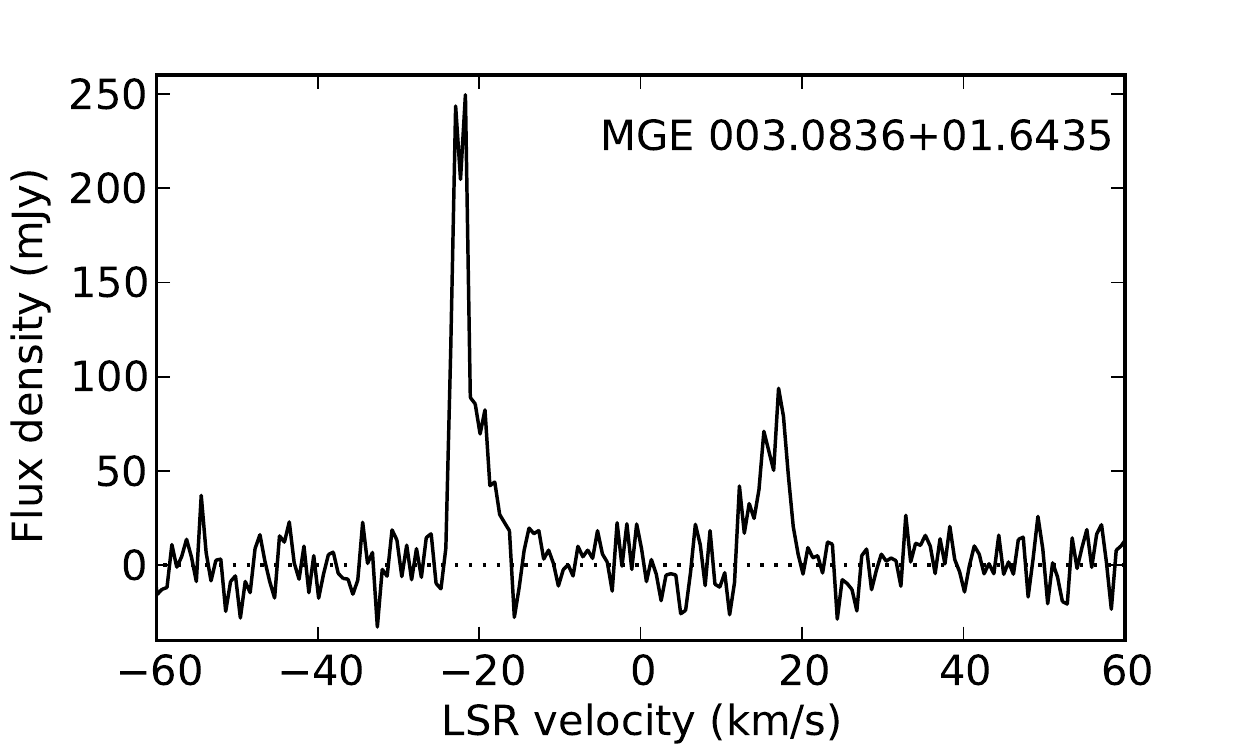}\hspace{-0.25cm}
\includegraphics[width=6cm]{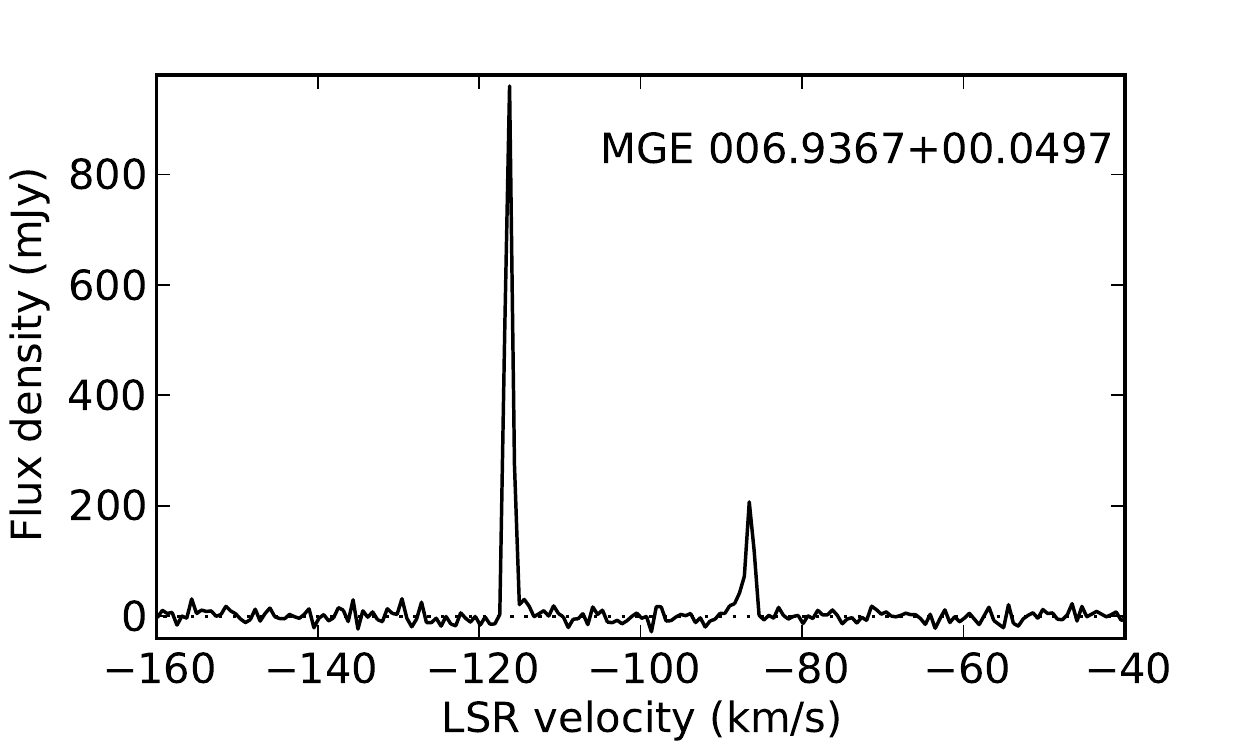}
\includegraphics[width=6cm]{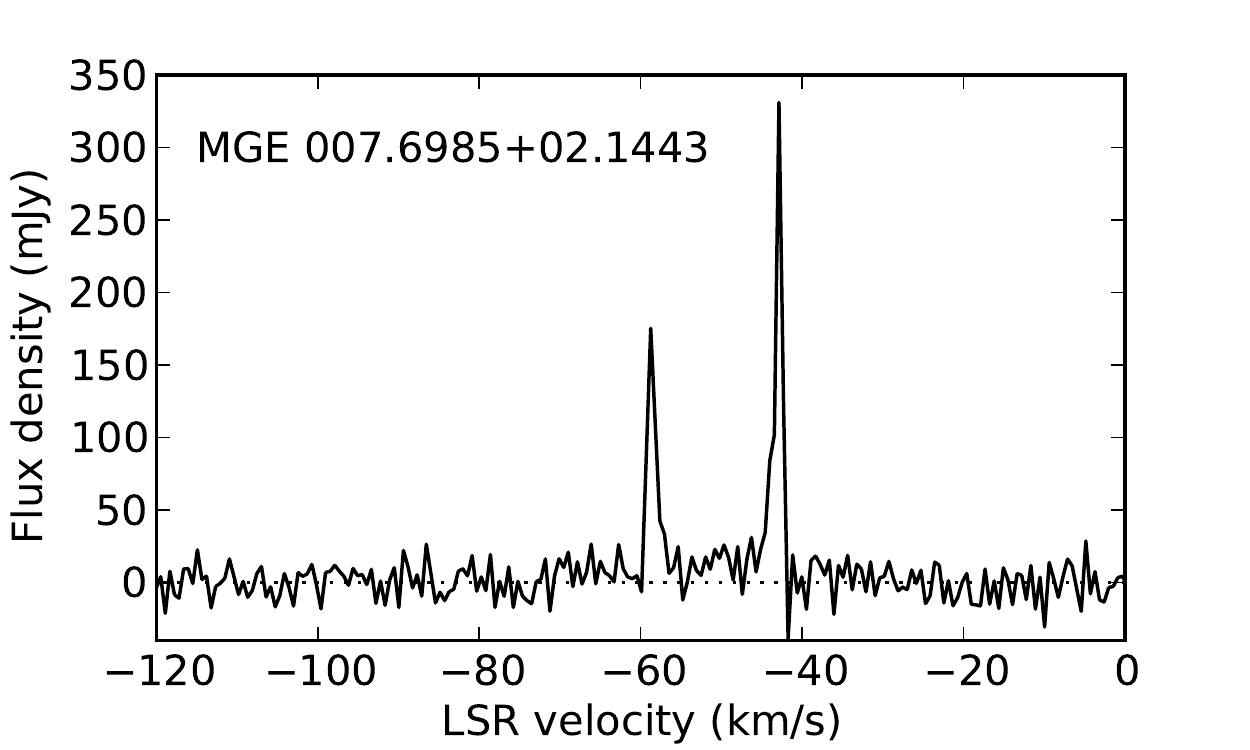}\hspace{-0.25cm}
\includegraphics[width=6cm]{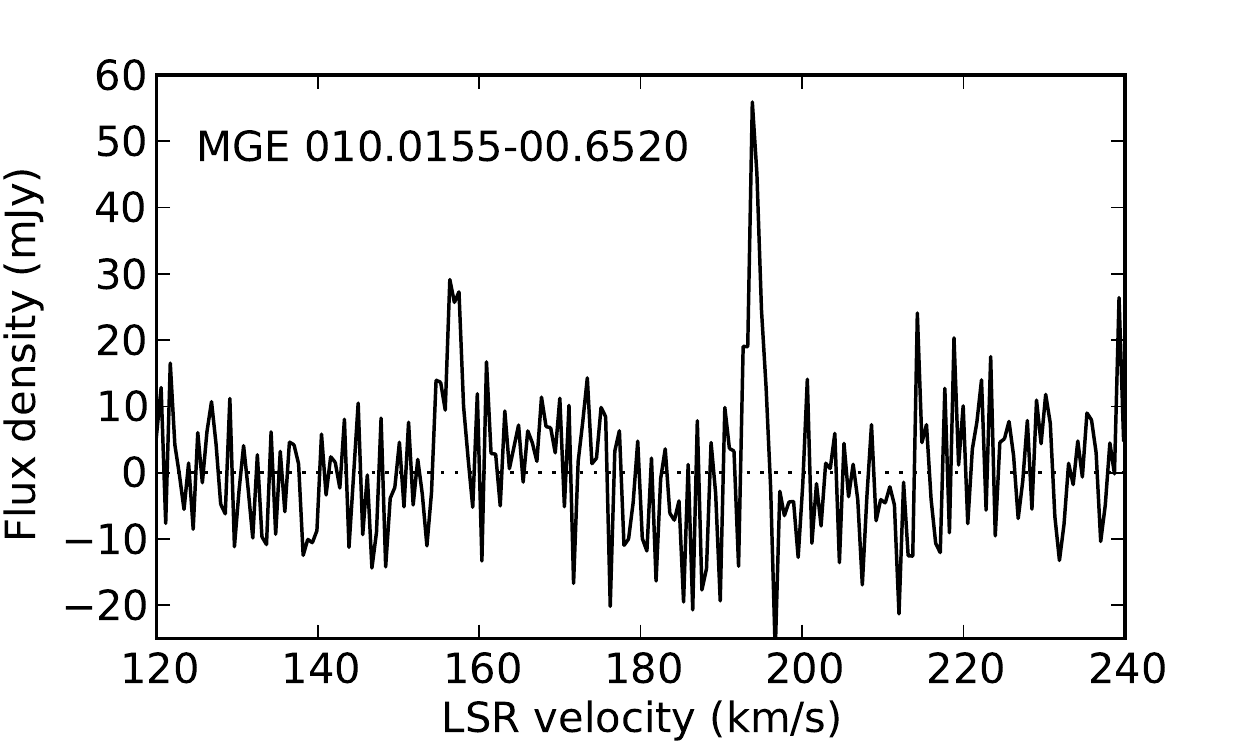}\hspace{-0.25cm}
\includegraphics[width=6cm]{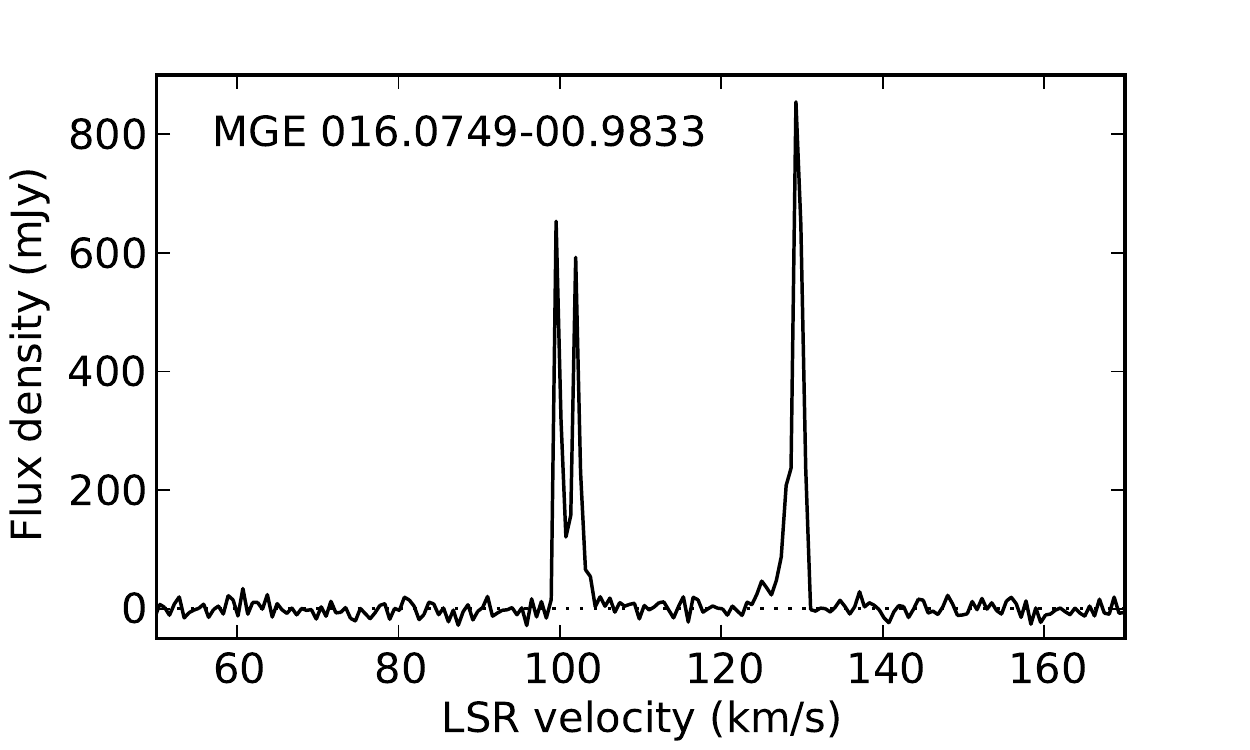}
\includegraphics[width=6cm]{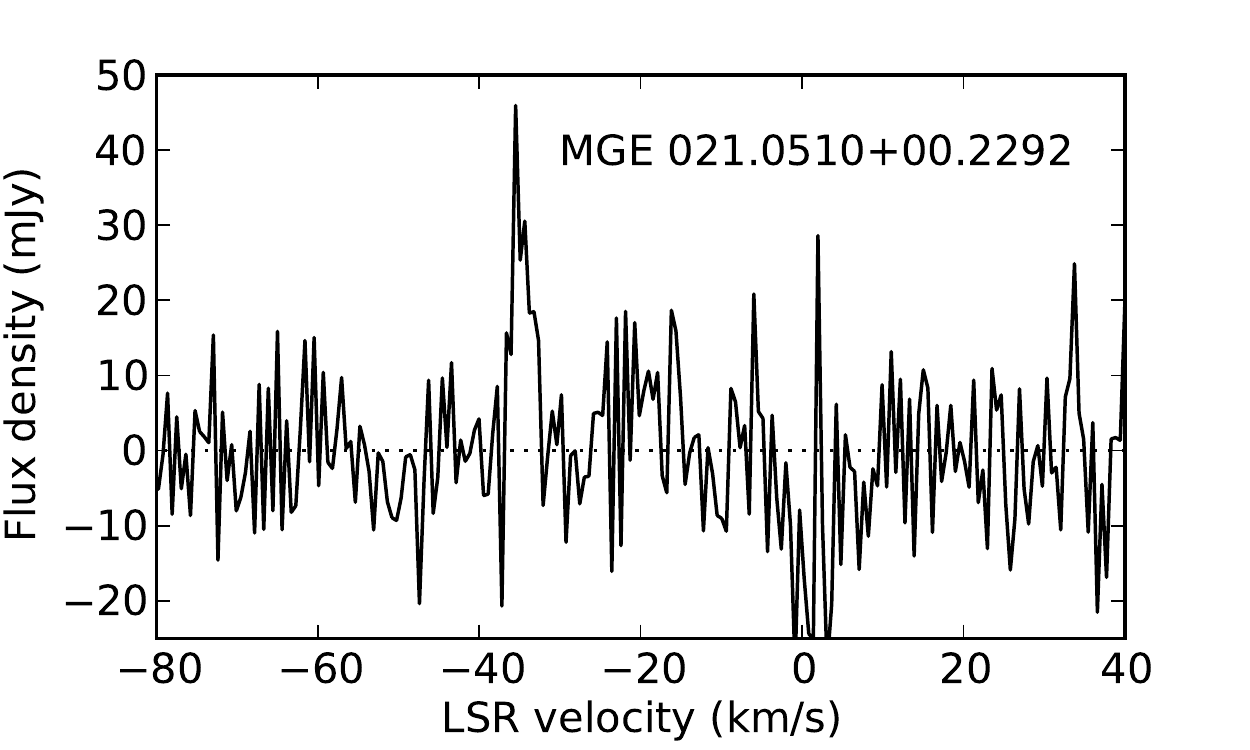}\hspace{-0.25cm}
\includegraphics[width=6cm]{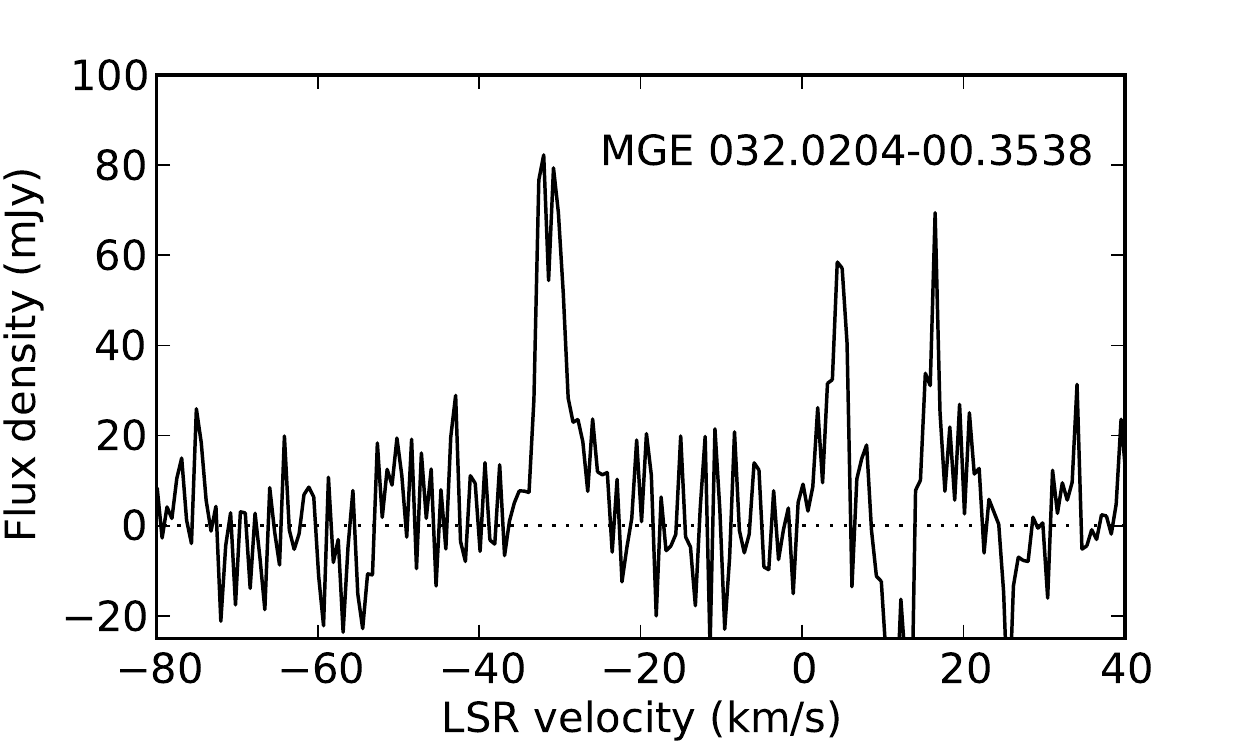}\hspace{-0.25cm}
\includegraphics[width=6cm]{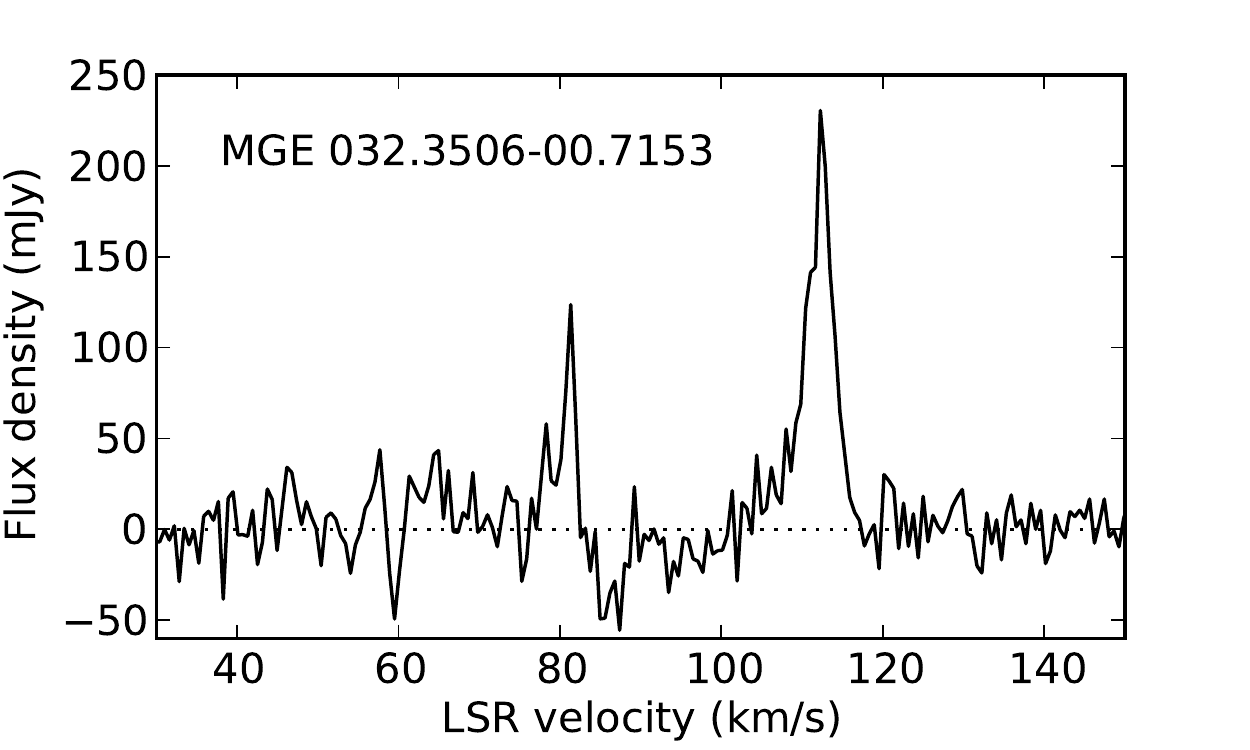}
\includegraphics[width=6cm]{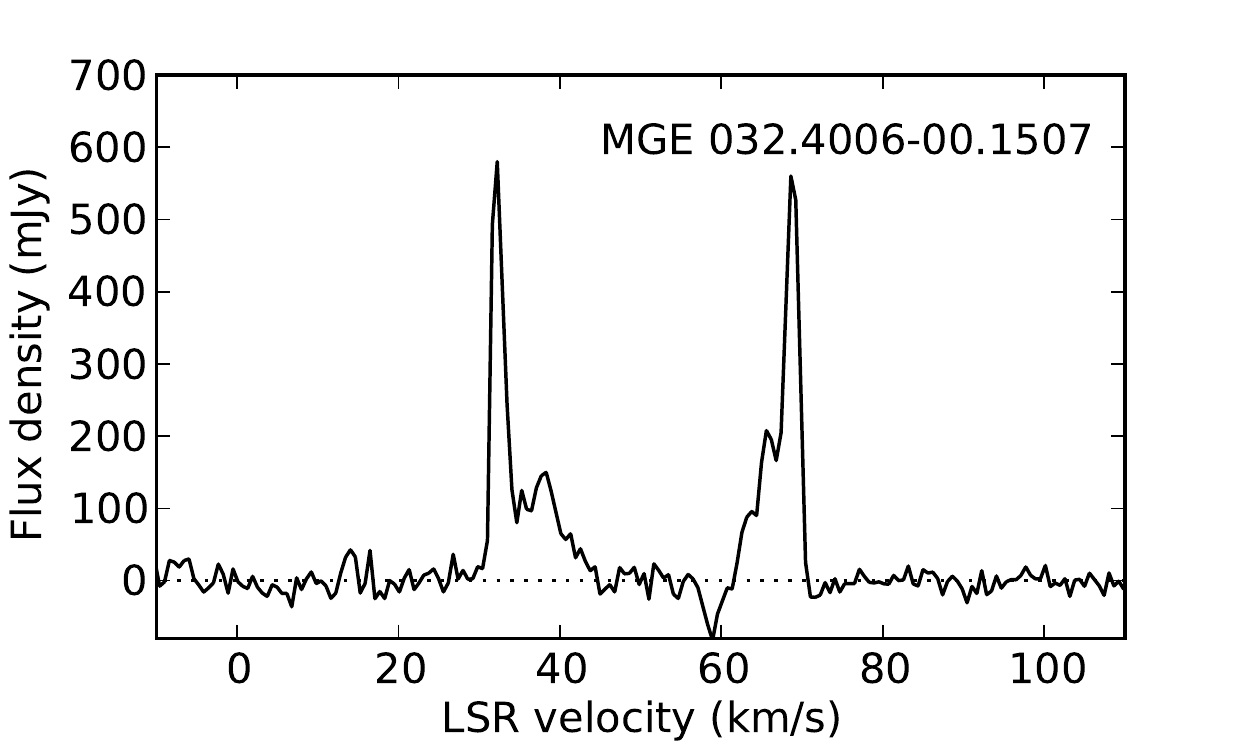}\hspace{-0.25cm}
\includegraphics[width=6cm]{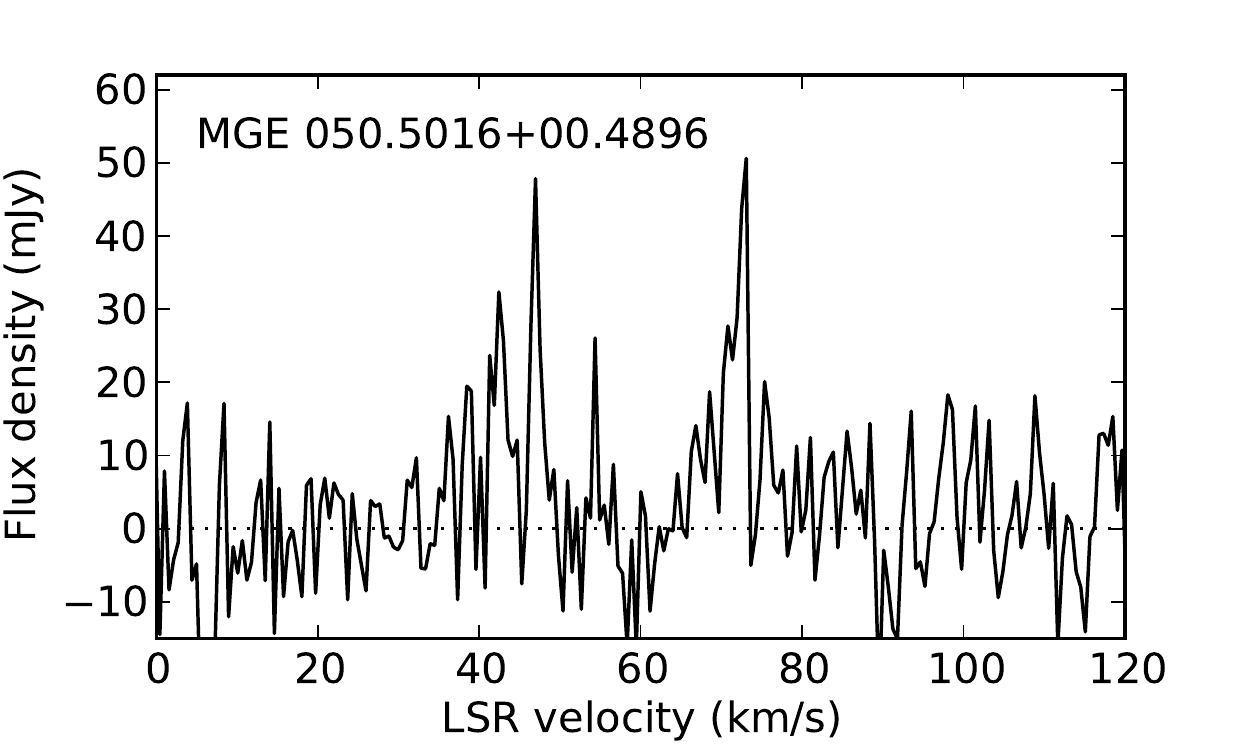}\hspace{-0.25cm}
\includegraphics[width=6cm]{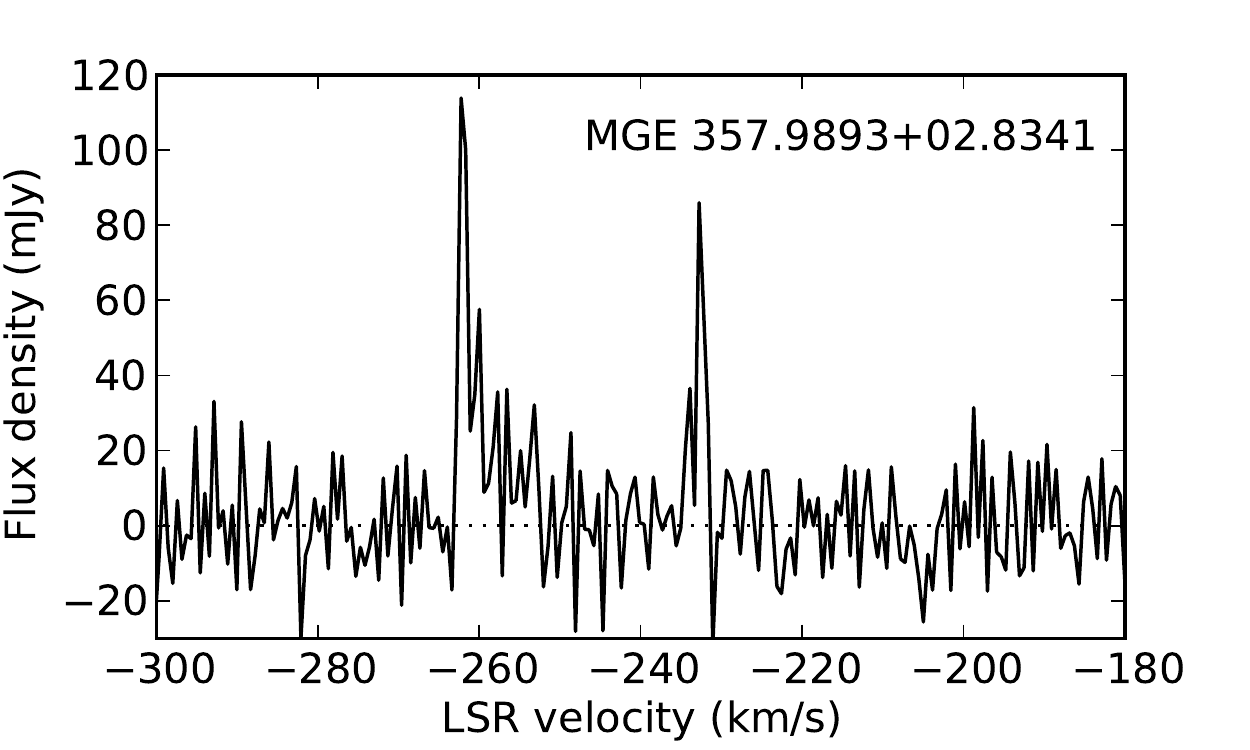}
\includegraphics[width=6cm]{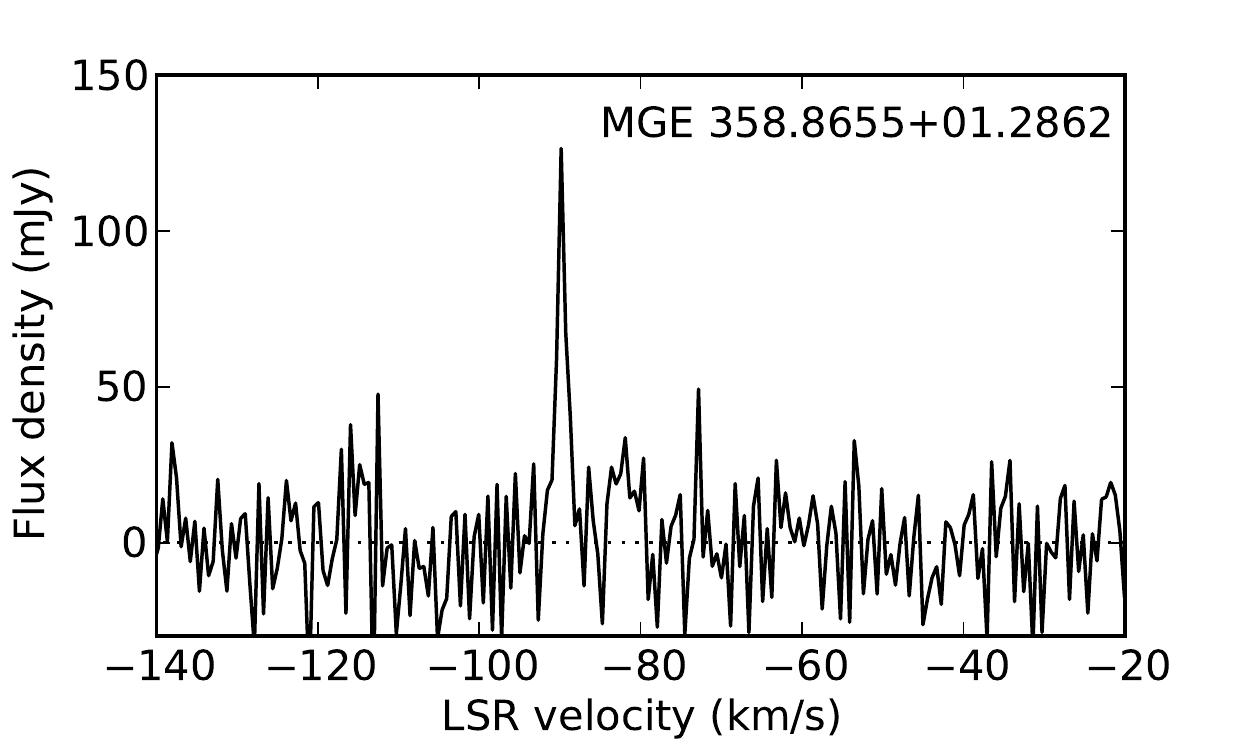}\hspace{-0.25cm}
\includegraphics[width=6cm]{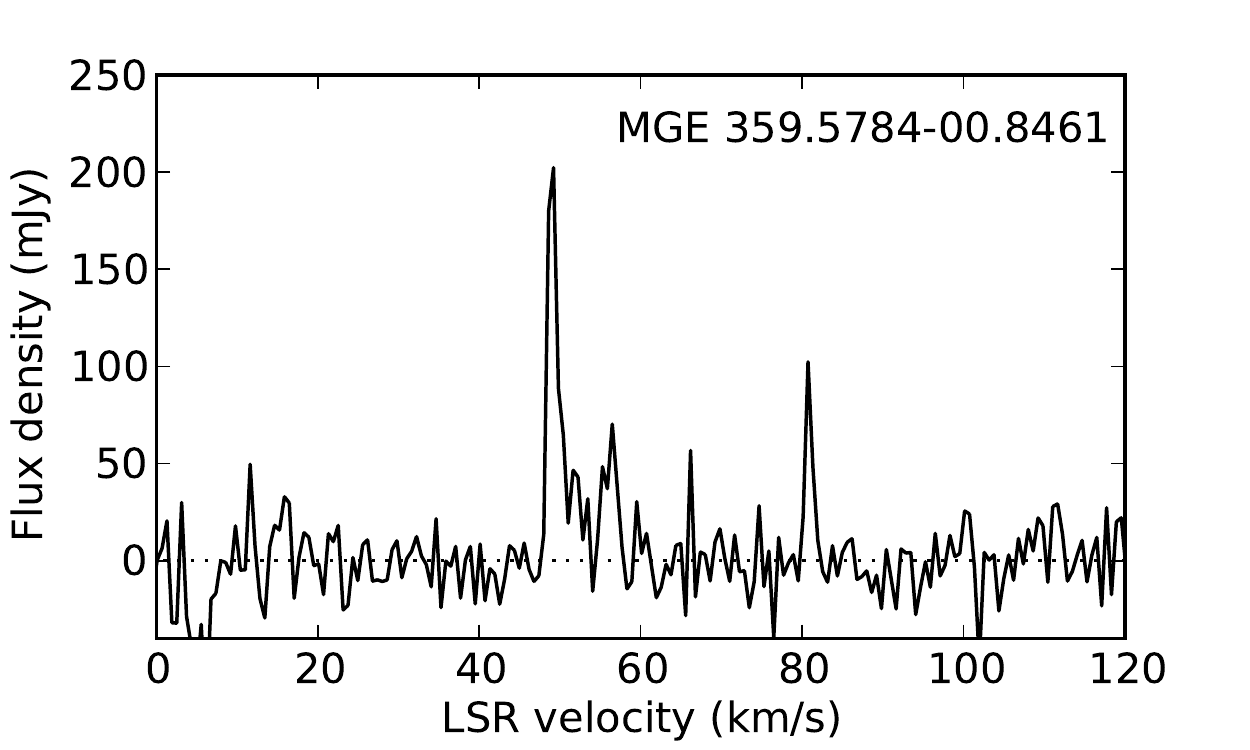}\hspace{-0.25cm}
\includegraphics[width=6cm]{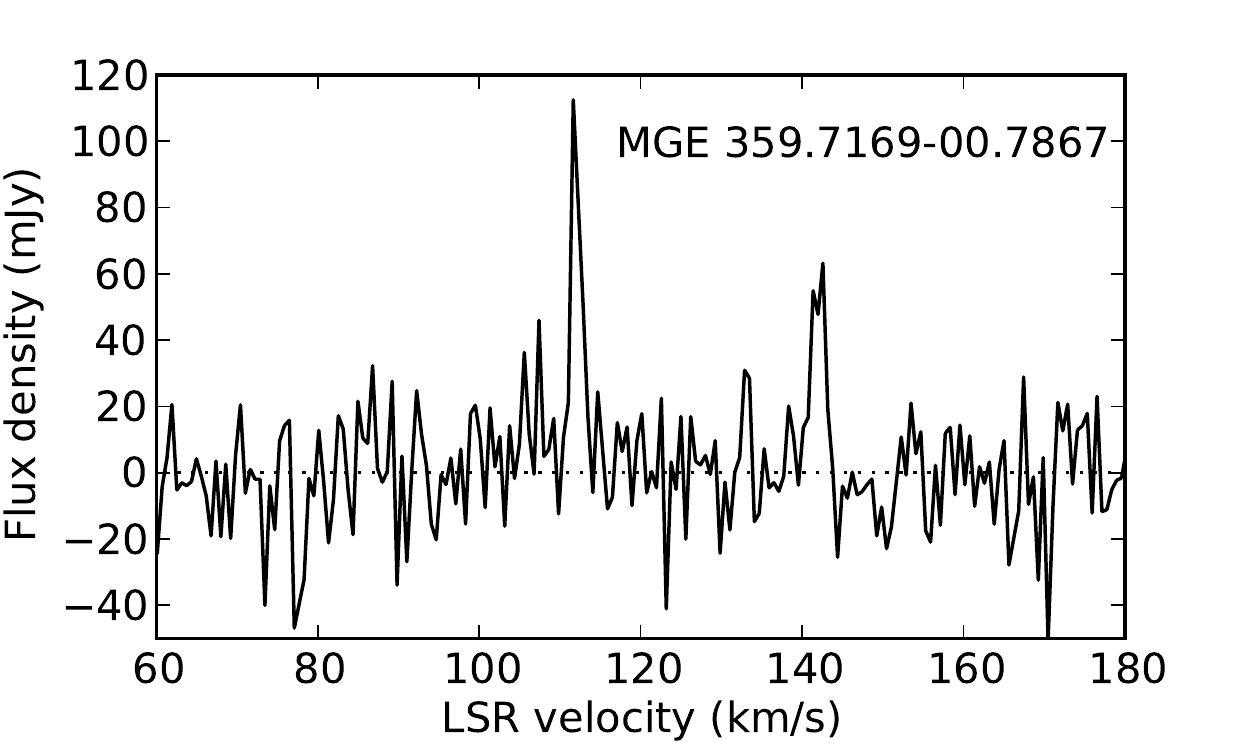}
\caption{Spectra at $1612\um{MHz}$ towards MBs with an uncertain match. Please notice that the velocity span is the same for all the spectra ($120\um{km\ s}^{-1}$) for an easier comparison of the D-type line separation. The two D-type lines of source MGE 000.4284+00.2292 fall inside a very wide ISM absorption line. For source MGE 001.2693+00.7174 two different spectra are given to better show the two different D-type lines observed in that direction.}
\label{fig:uncertain}
\end{center}

\end{figure*}

The 25 spectra with maser lines that do not have a catalogue identification might be referable to the MBs. To check if these matches are plausible we decided to produce datacubes from the aforementioned survey. \citet{Sevenster1997a} in fact did not produced images of their observations due to the extreme computational load required. They limited their search to an automatic method that was able to identify only very bright masers (the typical rms was $\sim\!30\um{mJy}$, but with significant variations across the field). A complete imaging of the survey is far beyond our computational possibility, therefore for each new detected maser we used the survey data to create small datacubes ($2\um{arcmin}$ across) centred in the MB positions. At the end of this process, we were able to discard 7 detections and to reasonably confirm one (Section \ref{sec:3337}). Due to the high noise, the matching of 17 spectra remained uncertain (Figure \ref{fig:uncertain}). In the Table \ref{tab:det} we report the confirmed and the uncertain matches. For the excluded matches see Appendix A.

\begin{table*}
\begin{center}
\begin{tabular}{ccrrrrrrccl}\hline
[MGE] & Type & \multicolumn{1}{c}{$v_b$} & \multicolumn{1}{c}{$S_b$} & \multicolumn{1}{c}{$v_r$} & \multicolumn{1}{c}{$S_r$} & \multicolumn{1}{c}{$\Delta v_\mathrm{peak}$} & \multicolumn{1}{c}{$\sigma$} & Morphology & Diameter & Notes\\
& & \multicolumn{1}{c}{($\mathrm{km\ s}^{-1}$)} & \multicolumn{1}{c}{(mJy)} & \multicolumn{1}{c}{($\mathrm{km\ s}^{-1}$)} & \multicolumn{1}{c}{(mJy)} & \multicolumn{1}{c}{($\mathrm{km\ s}^{-1}$)} & \multicolumn{1}{c}{(mJy)} & at $24\mic{m}$ & (arcsec)\\\hline
000.4284+00.2292 & D & 106.7    & 170.0 & 136.8    & 761.9 & 30.1 & 17.1 & 3b & 14\\
                 & D & 124.3    & 402.2 & 155.0    & 280.7 & 30.7 & 17.1\\
001.2693+00.7174 & D & $-120.5$ & 379.4 & $-93.2$  & 194.2 & 27.3 & 13.2 & 3d & 16\\
                 & D & $-0.5$   & 80.9  & 29.8     & 76.7  & 30.3 & 13.2\\
001.8248-00.9674 & D & $-183.5$ & 151.7 & $-151.4$ & 69.1  & 32.1 & 12.9 & 3a & 16\\
003.0836+01.6435 & D & $-21.7$  & 249.5 & 17.1     & 93.9  & 38.8 & 10.9 & 3b & 13\\
006.9367+00.0497 & D & $-116.2$ & 960.0 & $-86.6$  & 207.2 & 29.6 & 9.5  & 3b & 20\\
007.6985+02.1443 & D & $-59.7$  & 175.3 & $-42.9$  & 331.0 & 16.8 & 8.6  & 3e & 24\\
010.0155-00.6520 & D & 156.4    & 35.8  & 193.8    & 59.6  & 37.4 & 7.3  & 3e & 21\\
016.0749-00.9833 & D & 99.5     & 653.4 & 129.4    & 854.8 & 29.9 & 10.7 & 2b & 25\\
021.0510+00.2292 & S & $-35.5$  & 45.9  &          &       &      & 6.1  & 1a & 45\\
029.2228+00.5392 & D & 56.6     & 164.9 & 85.5     & 81.6  & 28.9 & 9.3  & 3b & 16 & \textit{Confirmed match}\\
032.0204-00.3538 & D & $-32.5$  & 82.3  & 4.4      & 58.54 & 36.9 & 9.6  & 1b & 66\\
032.3506-00.7153 & D & 81.3     & 123.6 & 112.3    & 230.5 & 31.0 & 11.6 & 2b & 23\\
032.4006-00.1507 & D & 32.2     & 580.0 & 68.6     & 560.0 & 36.7 & 10.8 & 2b & 33\\
050.5016+00.4896 & D & 47.0     & 47.8  & 73.1     & 50.6  & 26.1 & 14.3 & 2b & 28\\
357.9893+02.8341 & D & $-262.2$ & 113.9 & $-232.8$ & 86.0  & 29.4 & 8.8  & 3e & 21\\
358.8655+01.2862 & S & $-89.8$  & 126.4 &          &       &      & 11.1 & 3a & 15\\
359.5784-00.8461 & D & 49.2     & 202.3 & 80.7     & 102.3 & 31.5 & 13.5 & 3d & 26\\
359.7169-00.7867 & D & 111.7    & 112.5 & 142.6    & 63.2  & 30.9 & 13.7 & 3a & 16\\
\hline
\end{tabular}
\label{tab:det}
\end{center}
\caption{Confirmed and uncertain matches at $1612\um{MHz}$. For D-type spectra velocities, $v$, and flux densities, $S$, are given with $b$ and $r$ subscripts referring to blue and red component; $\Delta v_\mathrm{peak}$ is the peak separation and $\sigma$ is the standard deviation. For S-type spectra only velocity and flux density are given. Morphology code are: 1a = regular objects with central sources; 1b = irregular objects with central sources; 2b = irregular rings; 3a = flat disks; 3b = peaked disks; 3d = oblong disks; 3e = irregular disks. Morphology and diameter (at $24\mic{m}$) are after \citet{Mizuno2010}.}
\end{table*}

It is noteworthy to remember that about 15 years elapsed from Sevenster survey to our observations. During this amount of time OH maser lines could have significantly varied in intensity, and even appeared or disappeared. This explains why some of our maser line detections are not present in the ATCA/VLA OH $1612\um{MHz}$ catalogue or why we miss to detect some of known masers. This also means that even the discarded sources could potentially be new masers truly arising from MBs. The exception is the maser detected towards MGE 023.3894-00.8753 which has been located in the relative datacube at the position $\alpha=18:37:03.0$, $\delta=-08^\circ\,53'\,59''$, $46\um{arcsec}$ away from the pointing centre. 

\subsection{MGE 029.2228+00.5392}
\label{sec:3337}
\begin{figure}
\begin{center}
\includegraphics[width=\columnwidth]{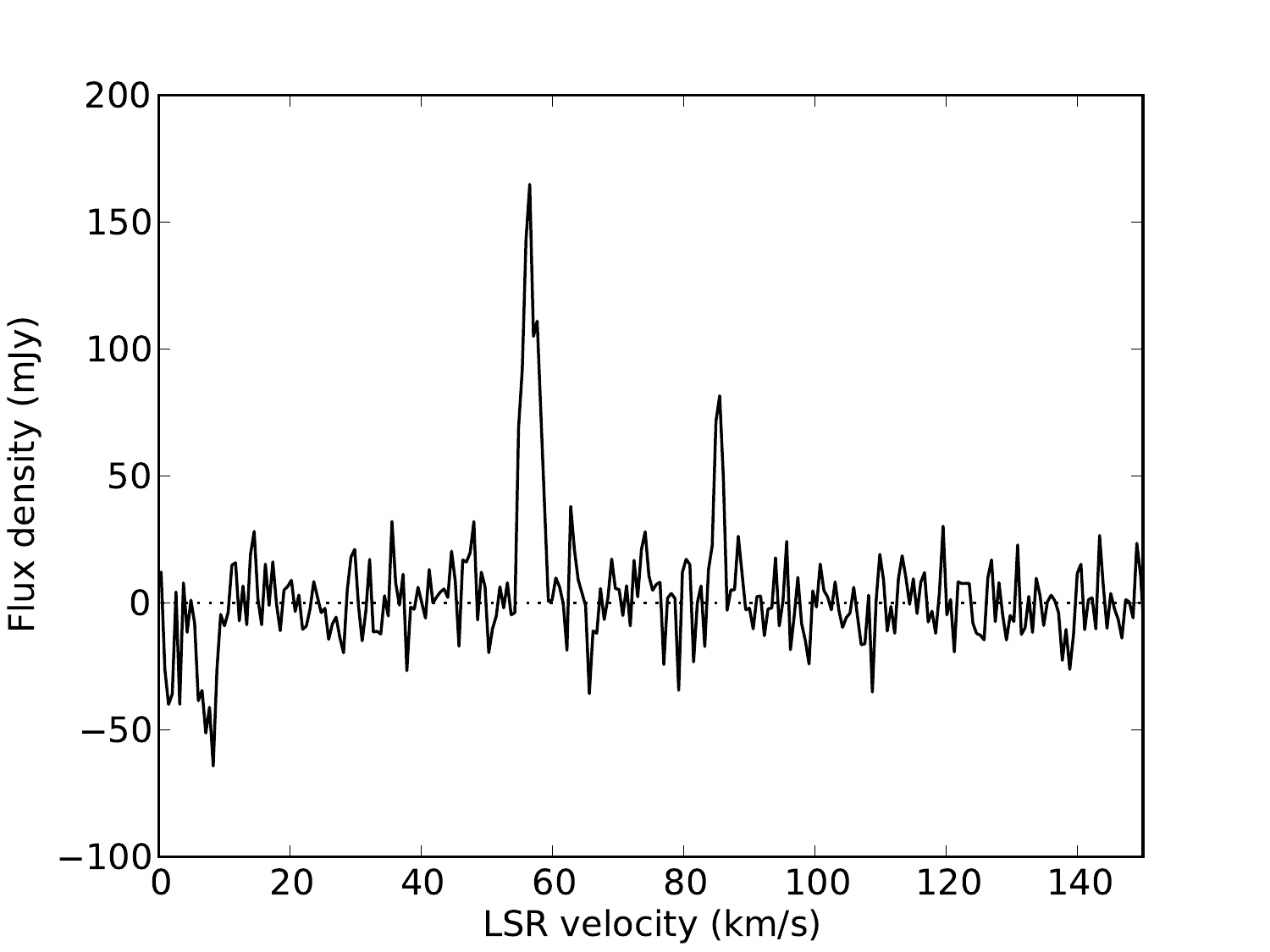}
\includegraphics[width=\columnwidth]{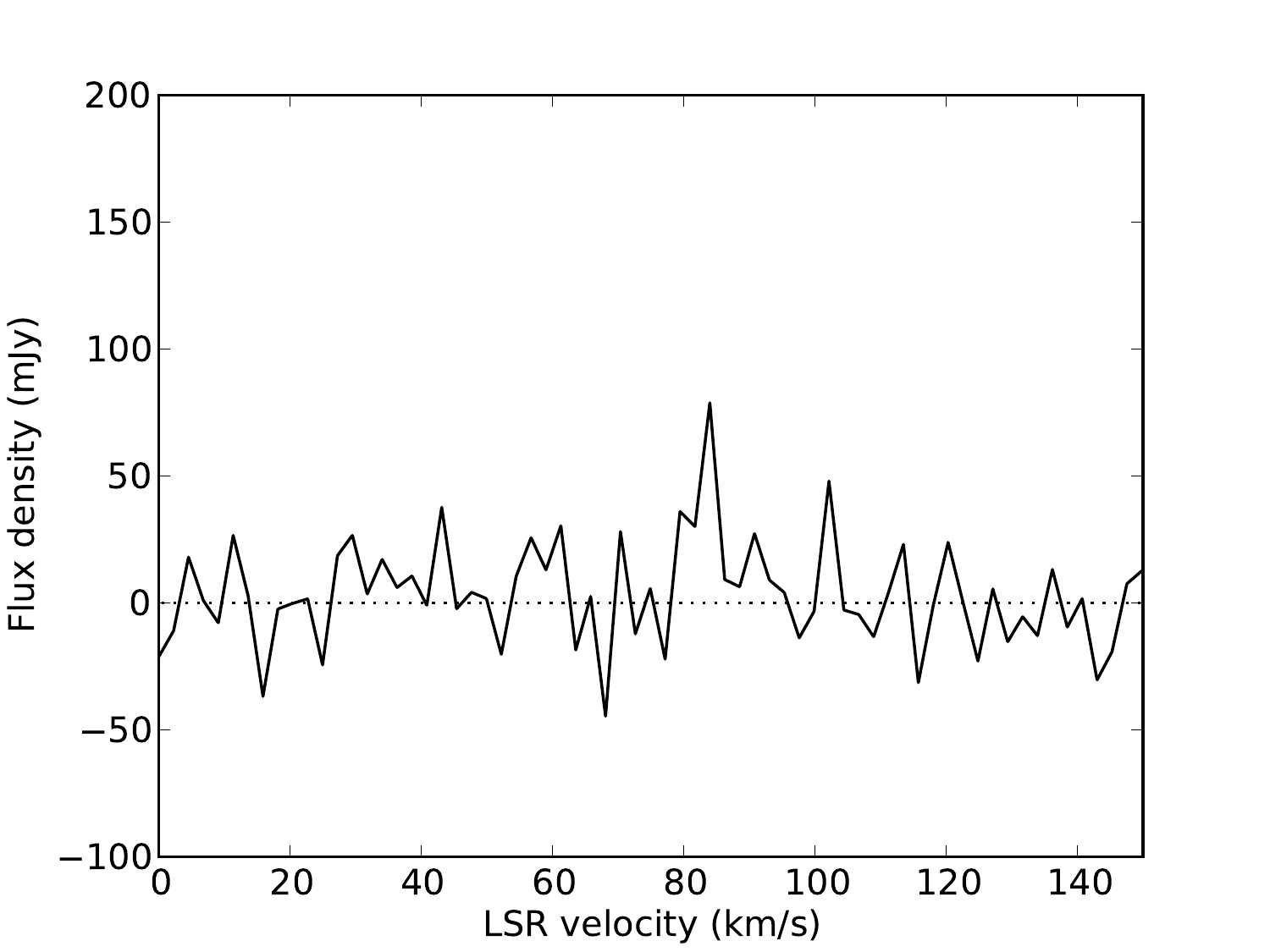}
\caption{Spectra around $1612\um{MHz}$ of the source MGE 029.2228+00.5392, from the GBT observations above and from the ATCA/VLA OH survey datacube below.}
\label{fig:3337_spec}
\end{center}
\end{figure}

The only spectrum with a maser line clearly associable with a MB is toward the source MGE 029.2228+00.5392. The GBT spectrum presents two prominent emission lines, which can be interpreted as one D-type emission or two S-types. Of these two lines only the redder one is unambiguously referable to this MB, according to our search in the relative datacube. In fact, this line is detected at a position only $\sim\!1\um{arcsec}$ away from the MB centroid position reported by \citet{Mizuno2010}, with the datacube pixel and beam size of, respectively, 0.5 and $1.5\um{arcsec}$. The position error associated with the datacube source is given by the beam size divided by the signal-to-noise ratio \citep{Purcell2013}, and it is equal to $\sim\!0.35\um{arcsec}$. This slightly displacement of the datacube source from the MB centroid is consistent with an OH maser origin a few thousands of AU (in projection) from the central star, while the CSE (with an angular radius of $8\um{arcsec}$) can extended up to tens of thousands AU. The density of known Galactic OH maser sources is about 1.4 per square degree \citep{Sevenster2001}. In our spectra we found that only about 45 percent of OH maser sources were alrady catalogued, so we could assume a density of 3 source per square degree with our sensitivity. The probability that such a source lies by chance within $1\um{arcsec}$ from the centroid of one of the 428 MBs is given by the total number of MBs multiplied by the area of a circle with 1-arcsec radius multiplied by the OH maser source density expected with our sensitivity. This number is about $10^{-7}$.
The line height is $78.9\pm18.3\um{mJy}$ ($4.3\sigma$) and is located at a velocity of $83.9\um{km\ s}^{-1}$. For comparison, from the GBT spectrum we derive a height of $81.6\pm9.3\um{mJy}$ at a velocity of $85.5\um{km\ s}^{-1}$. The good agreement between both velocities and heights makes us even more confident about the reliability of this match. In Figure \ref{fig:3337_spec} we report the GBT spectrum and the datacube one, extracted as explained in the previous section. The absence of the second line could be explained in two different ways: a) the line arises from another source outside the imaged datacube field; b) the line arises from the same source but its intensity has significantly increased in the last years.

\begin{figure}
\begin{center}
\includegraphics[width=\columnwidth]{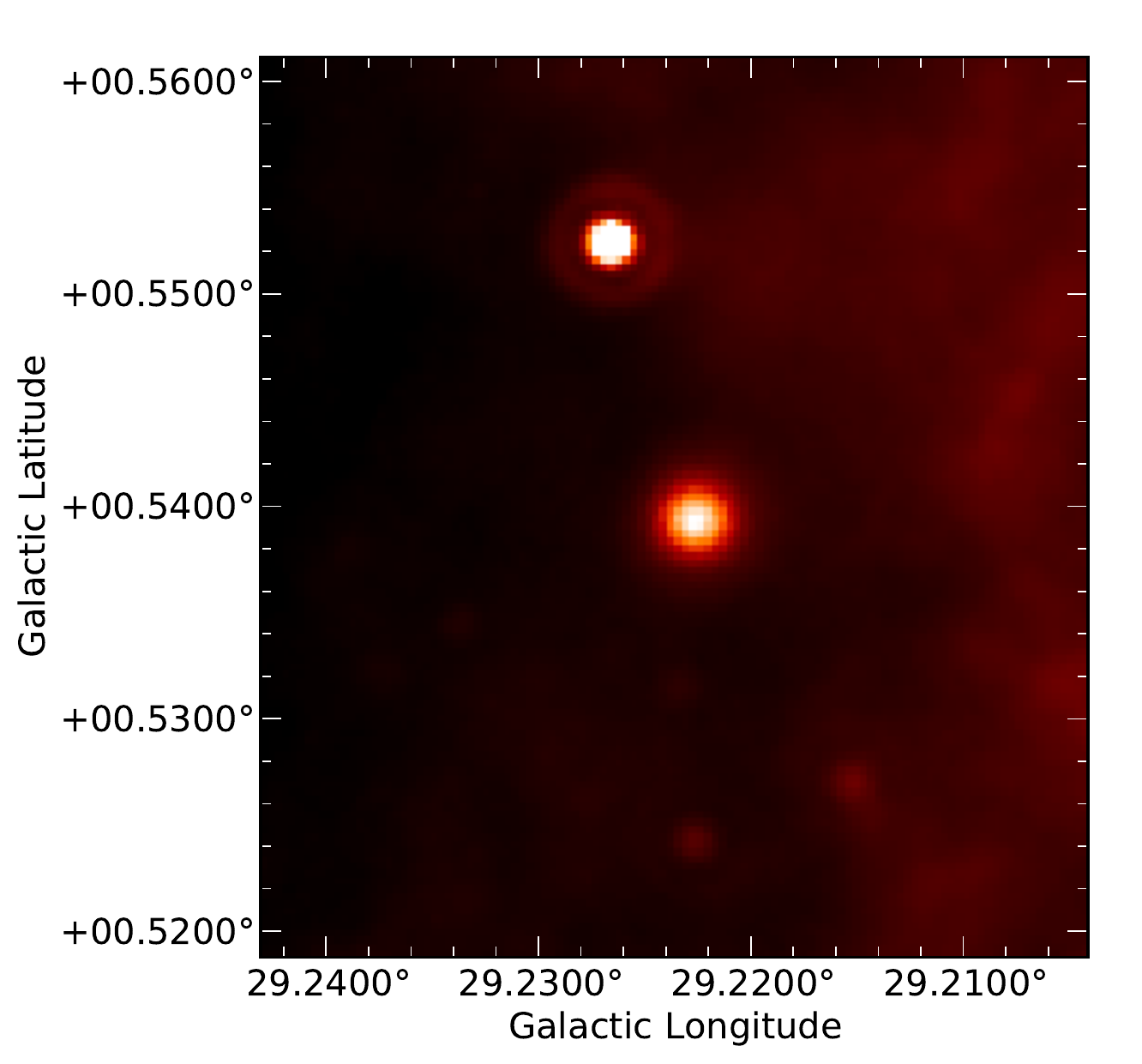}
\caption{MIPSGAL image at $24\mic{m}$ centred on the source MGE 029.2228+00.5392.}
\label{fig:3337_M24}
\end{center}
\end{figure}

The possible OH maser detection towards the bubble MGE 029.2228+00.5392 could be a fundamental clue for its identification. In the MIPSGAL tiles the source appears as a peaked `disk', the typical shape of PNe \citep{Nowak2014}, with a diameter of $16\um{arcsec}$ (see Figure \ref{fig:3337_M24}).

The flux density at $24\mic{m}$, found in literature, is $170\pm27\um{mJy}$ \citep{Mizuno2010}. Following the empirical relation between 24-$\umu$m and 6-cm flux density discussed in \citet{Ingallinera2014a}, for this source we could expect a flux density $\sim\!2\um{mJy}$ at $6\um{cm}$ if it were a radio thermal emitter. Unfortunately even the map with the highest sensitivity currently available (from the Galactic Plane Survey; \citealt{Helfand2006}) is too noisy to establish whether the radio emission is present or not. We can only give a flux density upper limit of $5.7\um{mJy}$ at $3\sigma$ level. Inspecting the GLIMPSE images, the source is not detected in any IRAC channel.

To the best of our knowledge, no systematic study has been conducted on this object, other than its discovery. Therefore the current information is too much limited to propose a clear classification. Its 24-$\umu$m morphology and the presence of the OH maser emission may suggest that this bubble could be a PPN or a very young PN.

\section{Summary and conclusions}
The classification of the MBs is still an open issue. Recent studies have confirmed the initial hypothesis that they are the CSE of different kinds of evolved stars. About a half of the currently identified MBs are PNe, while the others result related to massive stars. This percentage is however biased by the fact that the most studied MBs are those with the central sources, rarer than `disk'- or `ring'-type MBs usually classified as PNe \citep{Nowak2014}. Once a complete classification will have been reached, the PN number will be likely around 75 percent.

In this paper we have analysed the possibility of the presence of PPNe among the MBs. This search was driven by the importance of this evolutionary stage for the comprehension of the last phases of low- and intermediate-mass stars and their role in the chemical evolution of our galaxy. Furthermore some examples of known PPNe show IR morphologies very similar to MBs and one MB has already been classified as a PPN. Indeed known examples of post-AGB stars in MIPSGAL show that these stars usually appear at $24\mic{m}$ as bright point sources, with no evidence of a nebula either at shorter wavelengths. We argue that the typical shape for a potential PPN among the MBs would be a `disk'.

To search for possible PPNe we conducted radio observation searching for OH masers. We were able to confirm the presence of maser emission in only one of the 169 MBs we observed, while other 17 spectra show emission that still could be associated with MBs but whose matching is uncertain. On the other hand, 19 spectra show maser lines identified with field sources other than the targeted MBs and other 7 matches have been excluded through the comparison with interferometer data. If we do not consider the 17 uncertain matches, we have that among 27 spectra with maser lines only one is associated with a MB, that is about 4 percent. We can suppose that the same match rate can be applied to the 17 uncertain spectra, resulting that we can reasonably expect no more than one other positive match from them. Since our initial sample of 169 MBs is quite representative of the entire MB set (428 sources) we can state that the fraction of PPNe among the MBs is expected to be between 0.5 and 1.5 percent. We therefore conclude that PPNe, though present, could be a rare kind of evolved star among the MBs, very likely also because of their relative short lifetime with respect to the other evolved stars constituting the MBs.


\section*{Acknowledgements}
This work is based on observations made with the Green Bank Telescope and the Very Large Array of the National Radio Astronomy Observatory, a facility of the National Science Foundation operated under cooperative agreement by Associated Universities Inc.. Archive search made use of the SIMBAD database and the VizieR catalogue access tool, operated by the Centre de Donn\'ees astronomique de Strasbourg. The Digitized Sky Surveys were produced at the Space Telescope Science Institute under U.S. Government grant NAG W-2166. The images of these surveys are based on photographic data obtained using the Oschin Schmidt Telescope on Palomar Mountain and the UK Schmidt Telescope. The plates were processed into the present compressed digital form with the permission of these institutions.

\appendix

\section{Maser catalogue}
\begin{figure*}
\begin{center}
\includegraphics[width=6cm]{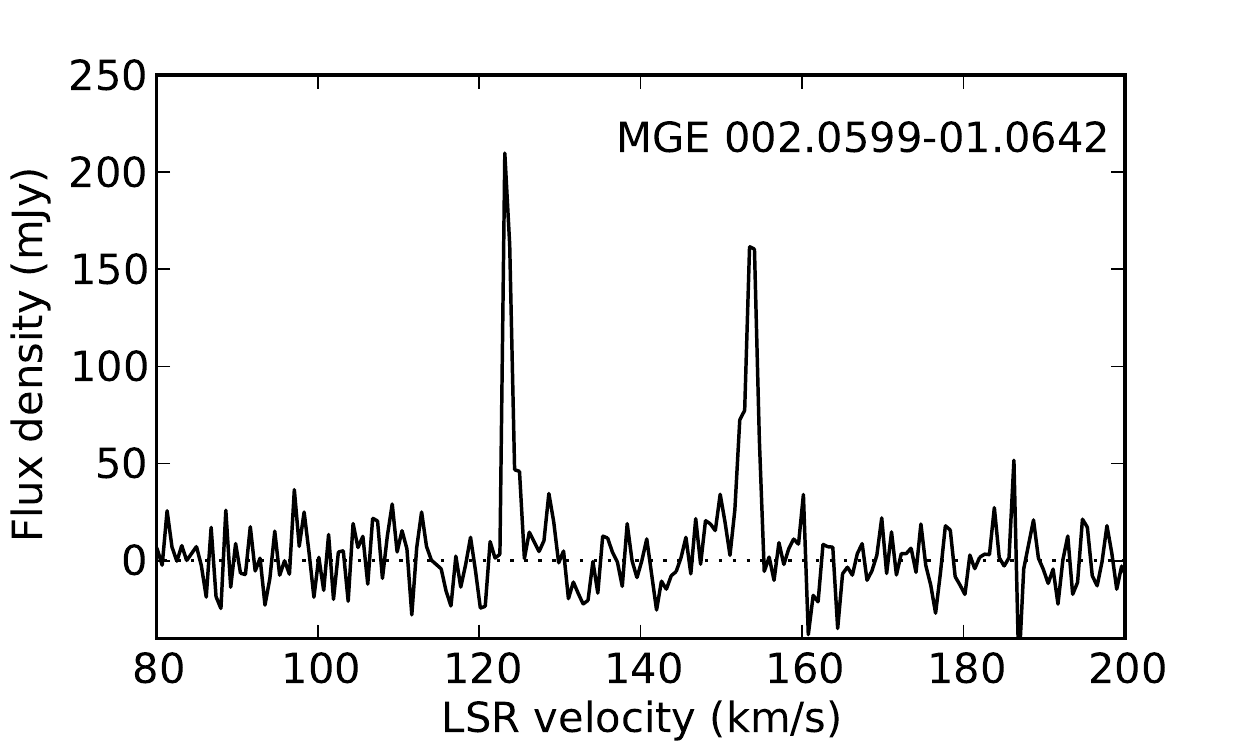}\hspace{-0.25cm}
\includegraphics[width=6cm]{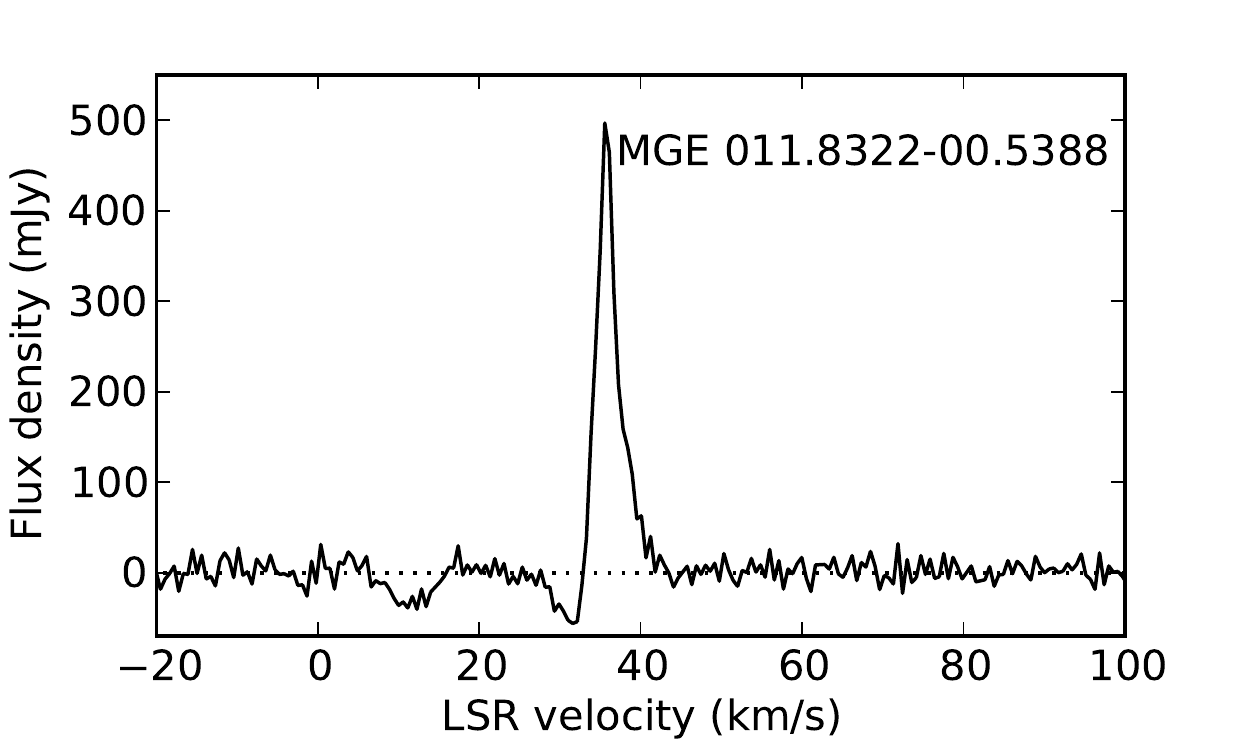}\hspace{-0.25cm}
\includegraphics[width=6cm]{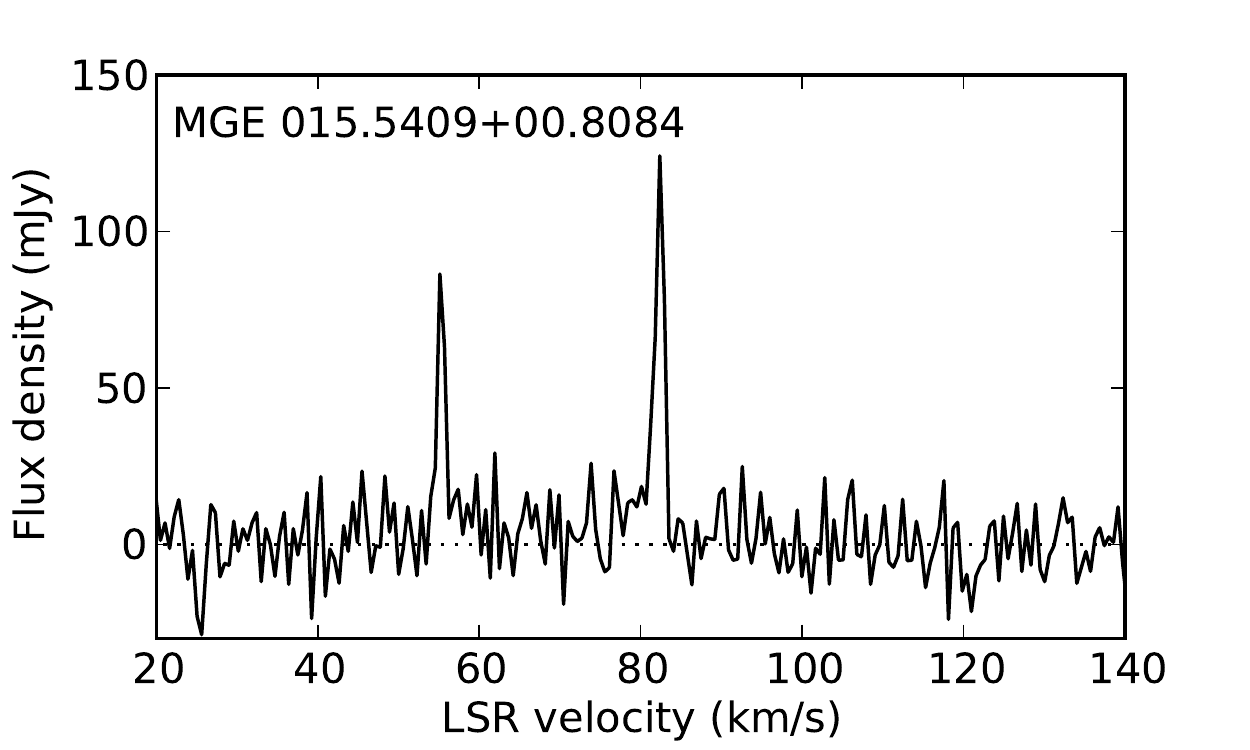}\hspace{-0.25cm}
\includegraphics[width=6cm]{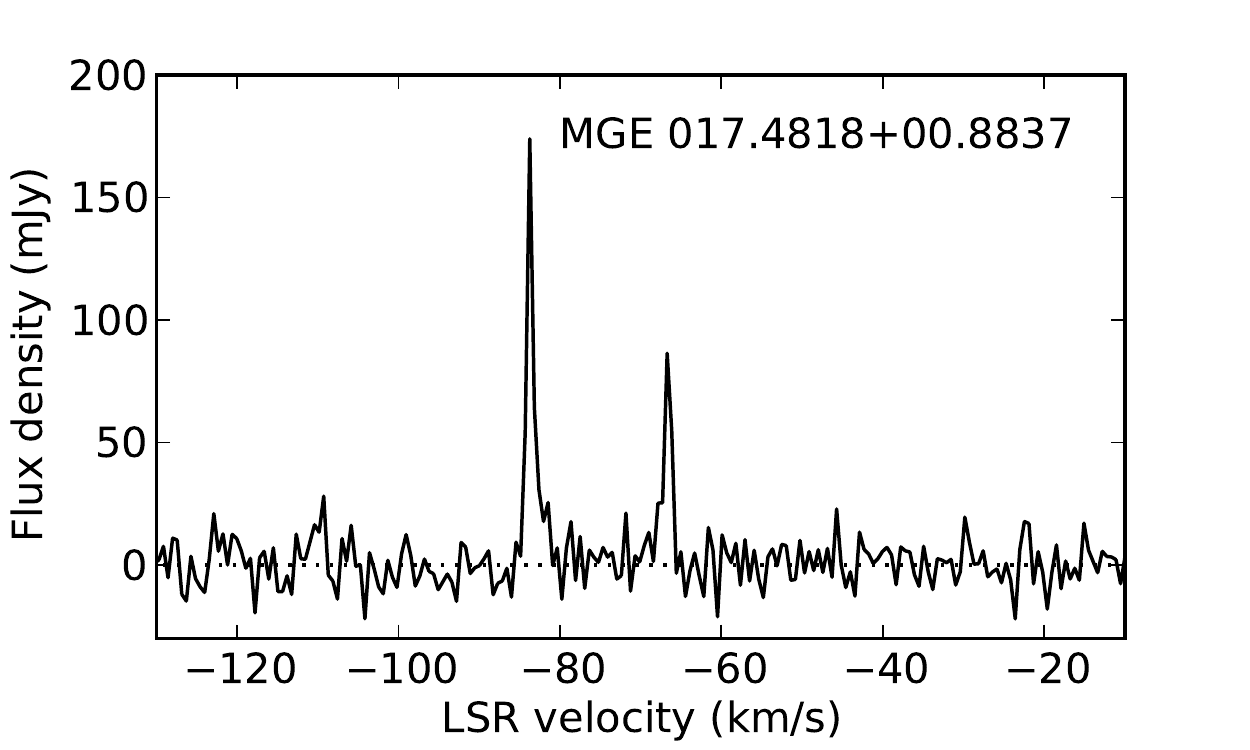}\hspace{-0.25cm}
\includegraphics[width=6cm]{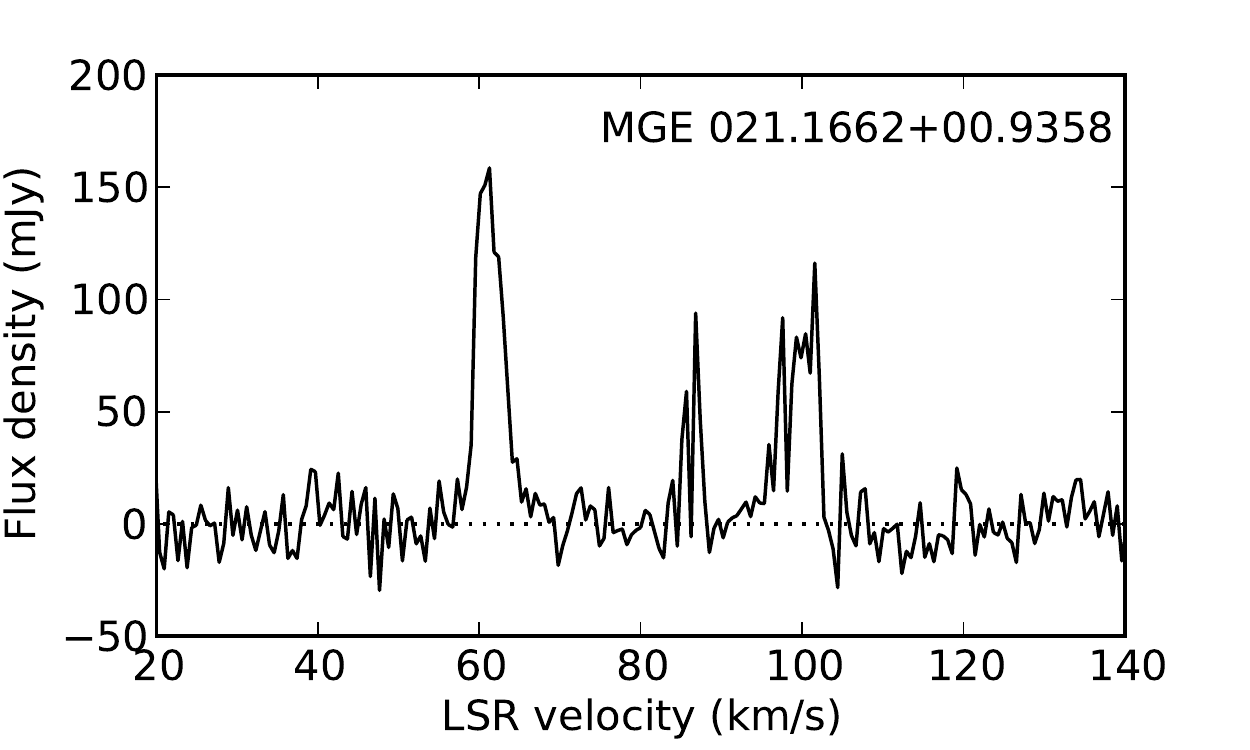}\hspace{-0.25cm}
\includegraphics[width=6cm]{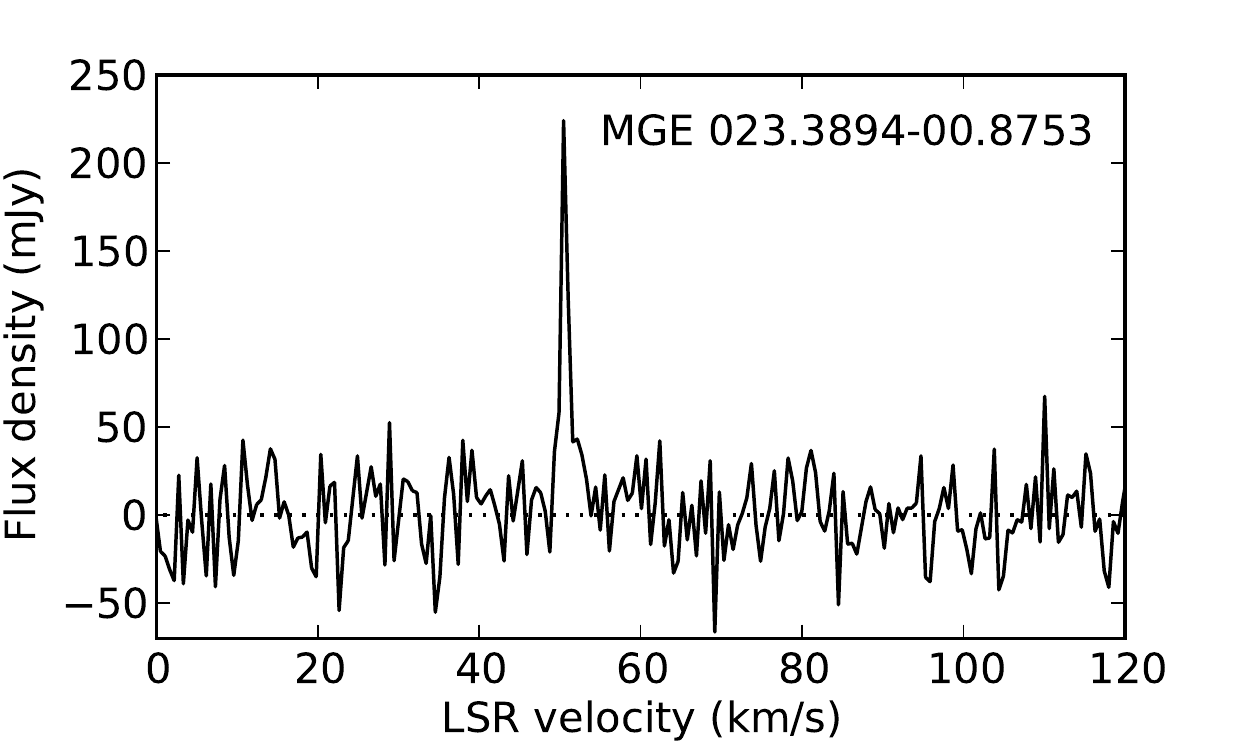}\hspace{-0.25cm}
\includegraphics[width=6cm]{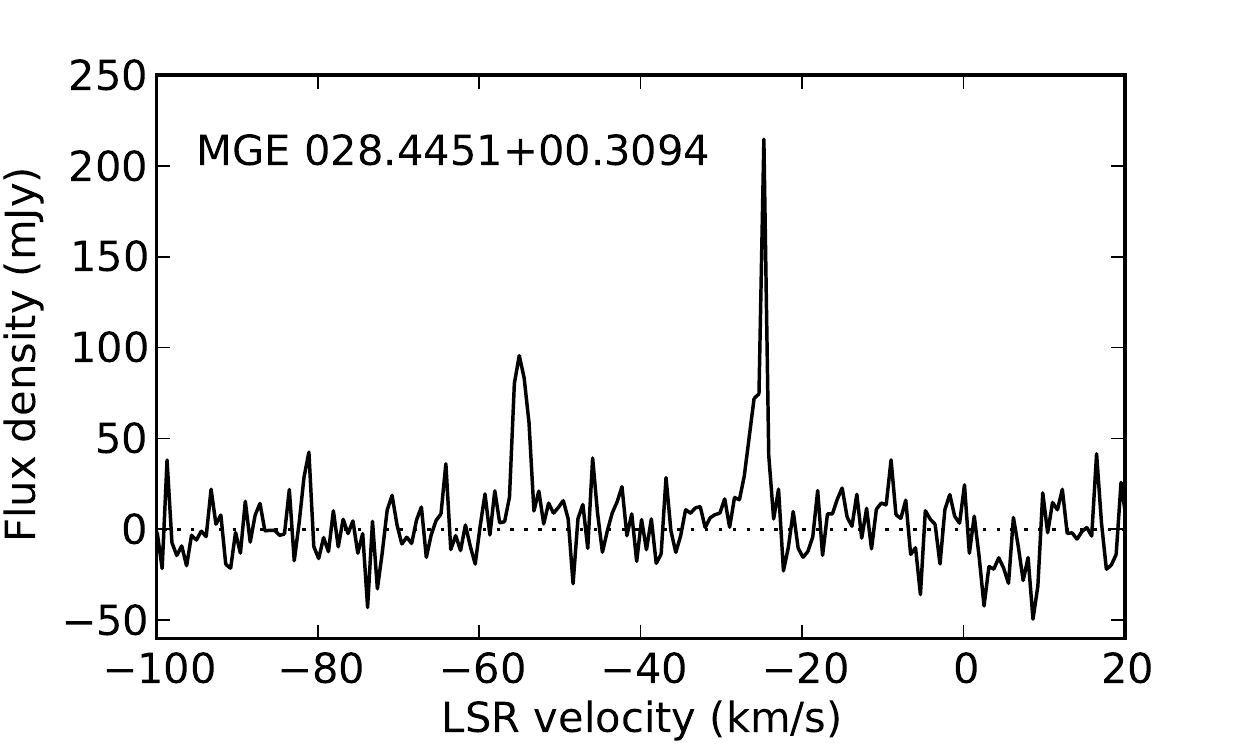}\hspace{-0.25cm}
\caption{Spectra at $1612\um{MHz}$ whose match has been excluded.}
\label{fig:excluded}
\end{center}

\end{figure*}

\begin{figure*}
\begin{center}
\includegraphics[width=7.5cm]{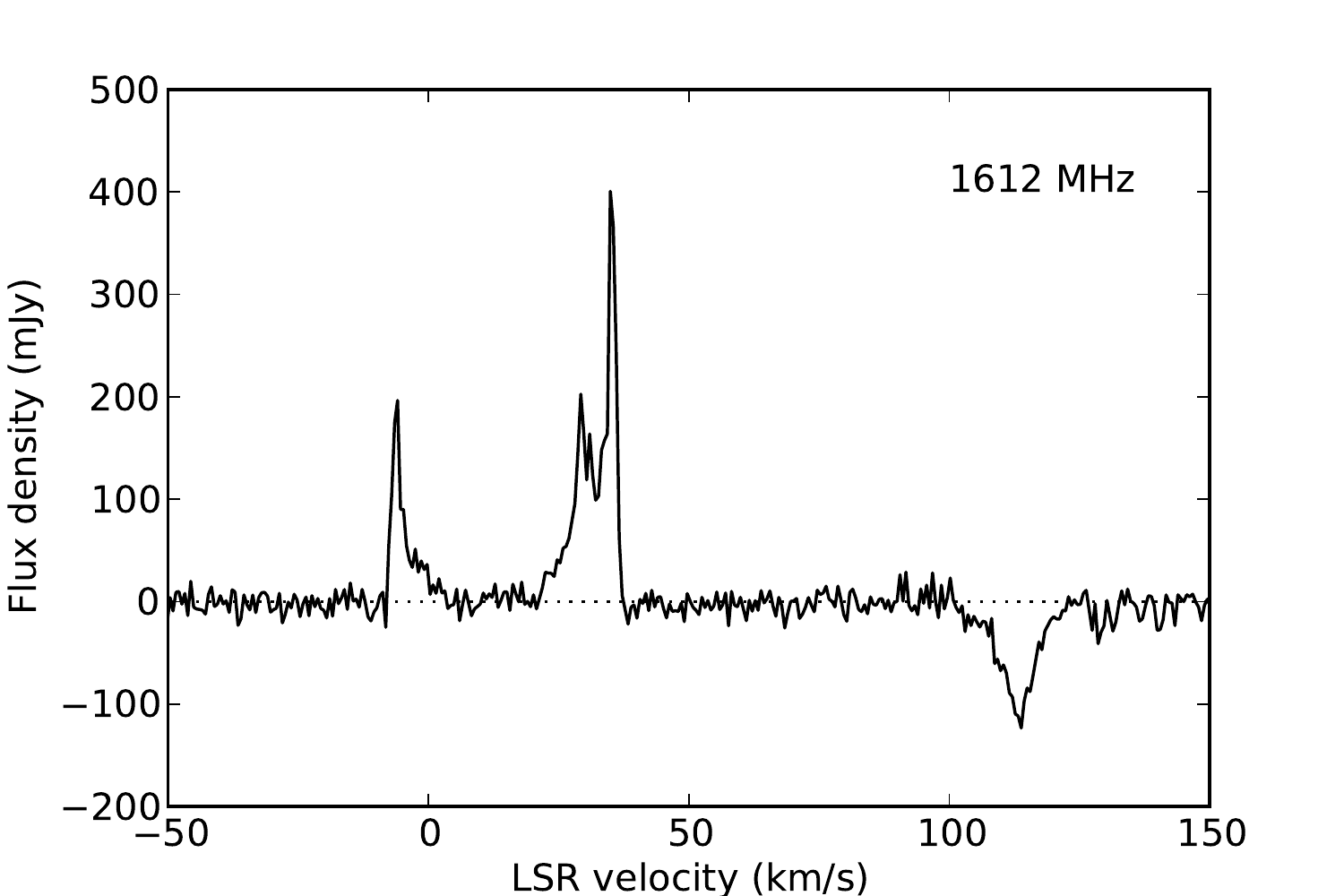}
\includegraphics[width=7.5cm]{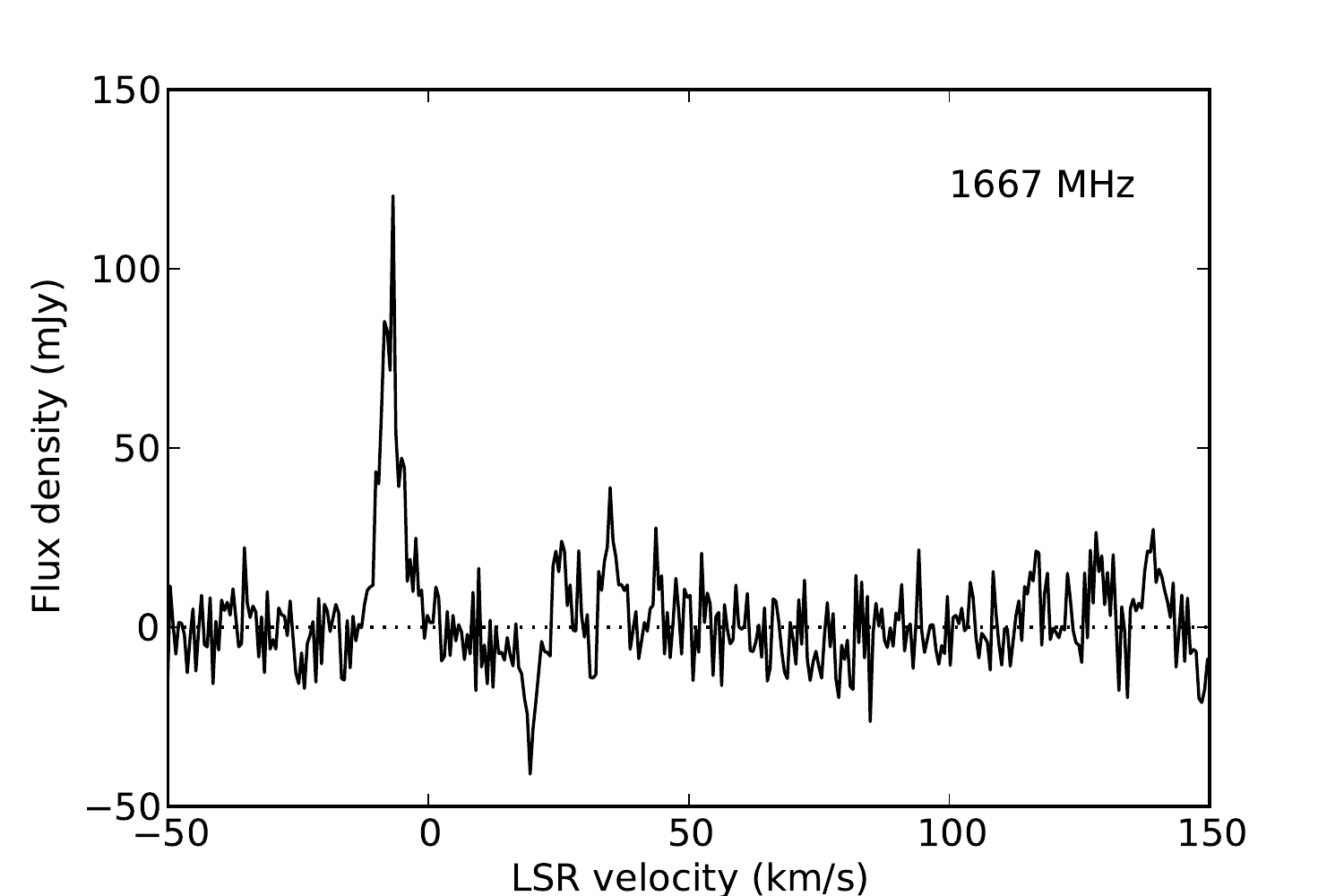}
\caption{Spectrum at 1612 and $1667\um{MHz}$ of OH016.247+00.174.}
\label{fig:1667}
\end{center}

\end{figure*}

In this appendix we list all the new maser detections that are not associable with MBs. In particular in Table \ref{tab:maser} we report all the maser at $1612\um{MHz}$ whose association with MBs has been excluded through the comparison with interferometer data. The relative spectra are showed in Figure \ref{fig:excluded}. In Table \ref{tab:maser_oS} we report the maser at $1667\um{MHz}$ associated with the known OH maser source OH016.247+00.174. Its spectrum at 1612 and $1667\um{MHz}$ is reported in Figure \ref{fig:1667}.

\begin{table*}
\begin{center}
\begin{tabular}{ccrrrrrr}\hline
[MGE] & Type & \multicolumn{1}{c}{$v_b$} & \multicolumn{1}{c}{$S_b$} & \multicolumn{1}{c}{$v_r$} & \multicolumn{1}{c}{$S_r$} & \multicolumn{1}{c}{$\Delta v_\mathrm{peak}$} & \multicolumn{1}{c}{$\sigma$}\\
& & \multicolumn{1}{c}{($\mathrm{km\ s}^{-1}$)} & \multicolumn{1}{c}{(mJy)} & \multicolumn{1}{c}{($\mathrm{km\ s}^{-1}$)} & \multicolumn{1}{c}{(mJy)} & \multicolumn{1}{c}{($\mathrm{km\ s}^{-1}$)} & \multicolumn{1}{c}{(mJy)}\\\hline
002.0599-01.0642 & D & 123.2   & 209.8 & 153.5   & 161.6 & 30.3 & 12.8\\
011.8322-00.5388 & S & 35.6    & 497.0 &         &       &      & 7.5\\
015.5409+00.8084 & D & 55.1    & 86.4  & 82.4    & 124.1 & 27.3 & 7.7\\
017.4818+00.8837 & D & $-83.7$ & 173.9 & $-66.7$ & 86.4  & 17.0 & 6.0\\
021.1662+00.9358 & D & 61.3    & 158.7 & 101.6   & 116.2 & 40.3 & 7.8\\
023.3894-00.8753 & S & 50.5    & 224.1 &         &       &      & 15.2\\
028.4451+00.3094 & D & $-55.0$ & 95.7  & $-24.7$ & 214.6 & 30.3 & 13.1\\
\hline
\end{tabular}
\label{tab:maser}
\end{center}
\caption{New detected masers at $1612\um{MHz}$, whose association with MBs has been excluded.}
\end{table*}

\begin{table*}
\begin{center}
\begin{tabular}{ccrrrrrrr}\hline
Frequency & Type & \multicolumn{1}{c}{$v_b$} & \multicolumn{1}{c}{$S_b$} & \multicolumn{1}{c}{$v_r$} & \multicolumn{1}{c}{$S_r$} & \multicolumn{1}{c}{$\Delta v_\mathrm{peak}$} & \multicolumn{1}{c}{$\sigma$}\\
(MHz) & & \multicolumn{1}{c}{($\mathrm{km\ s}^{-1}$)} & \multicolumn{1}{c}{(mJy)} & \multicolumn{1}{c}{($\mathrm{km\ s}^{-1}$)} & \multicolumn{1}{c}{(mJy)} & \multicolumn{1}{c}{($\mathrm{km\ s}^{-1}$)} & \multicolumn{1}{c}{(mJy)}\\\hline
1612 & D & $-5.9$ & 196.6 & 34.9 & 400.7 & 40.8 & 7.7\\
1667 & D & $-6.8$ & 120.4 & 34.9 & 38.9  & 41.7 & 6.5\\
\hline
\end{tabular}
\label{tab:maser_oS}
\end{center}
\caption{Masers associated with OH016.247+00.174. Flux densities are not corrected for GBT beam, so they are underestimated. It is possible to notice the evidence of the overshooting phenomenon \citep{Deacon2004} with the $\Delta v_\mathrm{peak}$ at $1667\um{MHz}$ slightly greater than at $1612\um{MHz}$.}
\end{table*}

\end{document}